# Performance augmentation in hybrid bionic systems: techniques and experiment.

**Bradly Alicea**
**MIND Lab, Michigan State University**

**Abstract:** Recent developments in hybrid biological-technological systems (hybrid bionic systems) has made clear the need for evaluating ergonomic fit in such systems, especially as users first become adjusted to using such systems. This training is accompanied by physiological adaptation, and can be thought of computationally as a relative degree of matching between prosthetic devices, physiology, and behavior. Achieving performance augmentation involves two features of performance: a specific form of learning, memory, and mechanotransduction called sensorimotor learning, and physiological adaptation to novel physical information imposed by the augmented environment of hybrid bionic systems. A method borrowed from environmental medicine involving perturbing the environment for a range of internal physiological conditions was used to induce sensorimotor learning and memory associated physiological changes. In addition, features of the adult phenotype were considered as a mitigator of learning-related adaptations. Using a series of statistical tests and techniques, the results demonstrate than three forms of regulation are at work related to morphological, neural, and muscular control. A discussion of the conceptual relationship between homeostasis and adaptation will then be discussed in addition to potential applications to performance augmentation strategies.

## 1.0 Introduction

This paper fills a specific gap in our knowledge regarding hybrid bionic systems (hybrid biological-technological systems): the theory and experiments presented here deal with the issue of physiological adaptation due to human-machine coupling through the use of a simple, noninvasive intervention. A secondary contribution lies in the presentation of evidence that is not usually brought to bear on the experimental problem. In doing so, readers can learn about the ultimate causes of adaptation and behavior. In this paper, the specific experimental test being used involves revealing the adaptability of human biological systems to environmental perturbations where the environment is manifest in the form of a human-machine system. In terms of application, this experiment may set up a series of design principles for hybrid systems such as brain-computer interfaces and limb prosthetics.

*1.1 Motivation.* This research provides a means to assess the degree of ergonomic matching (Bridger, 1995, Page 4) between environment and morphology under a range of conditions. Using the adaptive features of proprioception and movement behavior as a focal point allows for observation of changes in underlying physiological states and processes. In the research presented here, the goal is to selectively decouple a technological system from a physiological one, and then observe how they recouple over short time-scales. This recoupling occurs physiologically by means of an internal regulatory model and morphological mechanisms. To make this motivation more universal, examples of matching can also be observed in physical hybrid systems. In attempting to synchronize a motor-driven mechanical pendulum with a virtual animation of a pendulum, Gintautas and Hubler (2008) were able to distinguish between two dynamical system states: dual reality and mixed reality. Mixed reality states were defined as both pendulums being in-phase for extended periods of time, while dual reality states constituted a lesser degree of synchrony.

## 2.0 Literature Review

A computational neurobiological framework can be used to explain the relevance of concepts such as human-machine coupling, in addition to defining an internal model which plays a regulatory role for physiological adaptation and behavior during human-machine interaction. The first subsection introduces the idea of coupling between humans and machines. The second subsection introduces the concept of mapped physiological output as it relates to the connection between physiological regulation and virtual environments. Taken together, they provide rationale for the theoretical constructs of performance level and homeostasis. The third subsection introduces the concept of homeostasis[1] (see Box 1) as it relates to dynamic physiological regulation. The fourth subsection introduces the mechanisms behind physiological regulation. The fifth subsection introduces the area of sensorimotor learning and memory as it relates to sensory integration and control. Finally, the sixth subsection will introduce the consequences of skeletal muscle adaptation and standing morphological variation for hybrid bionic systems. All of these areas correspond with an internal model (Wolpert et.al, 1995) that treats the entire neuromuscular system as a regulatory mechanism.

**2.1 Computational neurobiology and the levels of analysis problem.** Shadmehr and Wise (2005, Chapter 1) have introduced a combination of method and theory called computational neurobiology. Using this approach, one can infer what transpires within neuromuscular systems by assuming that a set of physiological structures constitutes an internal model, which performs a regulatory function tightly linked to behavioral output. Sensorimotor behavior can be interpreted in terms of linear and nonlinear control theory (Jagacinski and Flach, 2003, Chapter 1; Tin and Poon, 2005; Fellous and Linster, 1998), which provides a mathematical basis for the underlying regulatory mechanisms of behavior. Another reason for this involves the focus on proprioception[2]. Recent work by Smith et.al (2001) with spinal cord injury victims has demonstrated that proprioceptive training[3] is the key to two-way communication in this cybernetic relationship between the human nervous system and machines. The brain, instead of being a black box, now becomes central to understanding what is going on at the behavioral level. In a like manner, phenomena at the cellular and molecular level are relevant to the behavioral level.

<div align="center"><b>Box 1. Definition of homeostasis and homeostasis vs. allostatis.</b></div>

In Sterling (2004), a distinction is made between homeostasis and allostasis. This distinction highlights the relevance of control theory to this regulatory context. Homeostasis holds a system constant by sensing deviation from a setpoint and using feedback to correct the error. This is an example of simple, closed-loop feedback. Allostasis is physiological change due to an environmental perturbation. In this context, the feedback loop is temporarily removed, resulting in open-loop control. Allostatic drive then can be defined as the movement from one physiological state to another via this mechanism.

*2.1.1 Definition of coupling between human and machine.* In recent years, the development and deployment of brain-machine interfaces for the recovery of function in quadruplegics (Nicolelis, 2003), smart prosthetics such as "living" heart valves (Flanagan and Pandit, 2003), and even the interface between intercochlear structures and cochlear implants (Johnson and Virgo, 2006, Pages 379-404 - see Box 2). One way to make the fit between biological systems and machines experimentally tractable is to focus on the effects of inertial feedback. In the context of performance augmentation, inertial feedback is a mixture of closed- and open-loop control that is explicitly related to inertial forces encountered during performance and physical control using a wearable or manual controller (Gillespie and Sovenyi, 2006; Huang et al. 1999). The effects of this coupling are also seen in the transformative[4] potential of a particular hybrid interface (Clark, 2003, Chapter 2). Smith (h2.0 Conference Webcast, 2007) has concluded that placing an interface as far away from the brain as possible allows for such an interface to take advantage of the plasticity exhibited by the entire nervous system (Fagg et.al, 2007).

---

[1] according to the traditional neurobiological/physiological definition, homeostasis involves the ability of a system to maintain energetic or ionic balance. More generally, this involves constancy in the internal state. In this document, homeostasis can be defined using an informational criterion.

[2] perception of force-related, inertial, and positional, and velocity-related stimuli that provide information about bodily movement, posture, and spatial orientation. This information can be unconscious (proprioceptive) or conscious (proprioceptive and kinesthetic). Afferent proprioceptive and kinesthetic information reaches the thalamus and cortex via the dorsocolumnar-medial lemniscus tract of the spinal cord, while afferent proprioceptive information also reaches the cerebellum via the spinocerebellar tracts.

[3] the repeated delivery of proprioceptive stimuli to optimize or degrade movement, alter kinesthetic awareness, and induce reactive and activity-dependent plasticity in the spinal cord and central nervous system.

**Box 2. Experiential aspects of hybrid bionic systems**

Hybrid bionic systems come in two versions: systems that involve a specialized interface that interfaces directly with the body, and technological systems that have an inadvertent effect on the internal physiological state. The first class of systems can be typified by the C-leg, a wearable prosthetic leg that is controlled by a series of microprocessors to aid in effortless walking. A related technology is the cheetah leg, which is specialized for running. It is of note that using the cheetah on one leg while having a normal leg on the other side might impede running speed, using a cheetah as a substitute for both legs might give the wearer an unfair advantage in competitive sport (McHugh, 2007). The reverse may actually be true in the case of the C-leg, as having one normal leg might allow the software in the C-leg to gauge the dynamics of movement behavior in the individual using it. In both cases, the proper application and perturbation of inertial feedback are critical to the discovering the optimal operation of these devices.

The second class of systems involves inadvertent or conditioned adaptation to a technological system used on a regular basis. One example is portable devices that affect function in sensorimotor systems. One example involves neurorehabilitation experiments in which postural stability is measured in relation to an auditory stimulus (see Chiari et.al, 2005). In experiments with a force platform and an iPod, continuous frequency and volume information acts to improve balance and postural control during standing. In a more naturalistic and dynamic setting, auditory or multisensory training might be used to improve muscular control. An example of this involves the deployment of wearable devices which continuously monitor balance at multiple points on the body and aid the user in correcting postural instabilities using auditory feedback (Brunelli et.al, 2006). On the other hand, such training might have deleterious effects depending on how it is implemented. In the case of phantom vibrations from cell-phone use, many people who use cell phones on a regular basis report experiencing vibrations similar to a ringtone in the place where they carry their phone in the absence of the device. This is similar to the "phantom limb" phenomenon, numerous example of which exist in the medical literature (for review, see Flor, 2002). In this and other cases, care must be taken in the application of perturbation to normal sensorimotor function. To avoid the negative consequences of augmentation, the pattern of perturbation might be modified to avoid constant of acute exposure to such a stimulus.

*2.1.2 Definition of Mapped Physiological Output.* The degree of coupling between the prosthetic device and nervous system function produces a measureable variable called mapped physiological output. One way to get at the relationship between adaptive forces acting upon internal model function[5] and resulting mapped physiological output is to understand how sensorimotor systems can be disembodied. This can be done using neural preparations where the output has been remapped to a machine. Reger et.al (2000) used the midbrain and brainstem of a lamprey (*Petromyzon marinus*) to control a Khepera robot[6]. In this demonstration, artificial photoreceptors on the robot body provide input to Muller cells in the brainstem, which was successful in integrating signals for use in navigation. Similar results have been achieved using hybrots (Potter et.al, 2004; Demarse et.al, 2001; Demarse and Dockendorf, 2005), which use neurons in culture to drive action in virtual and physical environments such as degrees of movement freedom in a flight simulator and robotic arms drawing on an easel. The bottom line of these studies is that such systems demonstrate a period of adaptation which is directly related to the compensatory ability of the nervous system to environmental variations that occur within a specific set of movement parameters (Shadmehr and Moussavi, 2000).

---

[4] in the context of transformative potential, this is the ability of an interface technology (e.g. implant, wearable device) to affect the state of physiological systems. Interfaces such as a wearable brace would have a greater transformative potential than would a cellular-scale implant because the former affects more physiological systems than the latter.

[5] two forms of plasticity that are operative in the brain and peripheral nervous system as a result of stimulation and adaptive behavior. Activity-dependent plasticity is defined by Wolpaw and Tennisen (2001) as driven by afferent inputs from the periphery and brain. This type of plasticity strictly causal, and contributes to the acquisition and maintainance of sensorimotor memory. Reactive plasticity is defined by Moller (2006, Page 271) as changes that occur as a side-effect of activity-dependent plasticity, usually in a functionally related part of the nervous system.

*2.1.3   Definition of physiological homeostasis.* Homeostasis can be defined as a series of tendencies and regulatory parameters that govern the stability of a particular system (Sterling and Eyer, 1988). According to Dworkin (1993, Pages 169-173), there are four properties of homeostasis that ultimately affect the adaptability of a hybrid system. The dynamic range of a system is determined by the degree to which a system adapts to perturbation. This may involve changes to either internal physiological processes or the capacity of sensory receptors to transduce information. The linearity of the system refers to how it responds to inputs, especially perturbations. Sensitivity refers to both the robustness[7] of the system and the magnitude of its compensatory capacity (see Box 3). Finally, the stability of a system refers to the capacity of a system to buffer noise.

It is the directional expression of this variation[8] towards a different homeostatic state that best characterizes allostatic drive in action (McEwen and Stellar, 1993). One example can be drawn from the sleep literature: Saper et.al (2005) propose that allostatic drive occurs when the sleep-wake cycle is interrupted due to the immediate demands of environment. Another example can be found in metabolic flexibility, which can allow for a response to environmental demands in healthy individuals (Blackstone et.al, 2005) but not in individuals exhibiting a disease state (Storlien et.al, 2004). Over the course of performance augmentation due to human-machine interaction, changes in these values are driven by both the need for a discrete physiological response. This adaptive ratchet mechanism drives an adaptive system in a particular direction. The ratchet is so called because adaptation is gradual, and driven by stepwise changes to the system's capacity (for an example from physical systems, see Freund and Poschel, 2000). In physiological systems, this may act to uncover latent physiological states, which are possible given the individual's genotype yet not expressed under normal conditions.

### Box 3. Definition of compensatory capacity.

Compensatory capacity is related to both homeostasis and allostasis. When homeostasis is maintained, the compensatory capacity relates to how much invariance a physiological system will exhibit while remaining in the same internal state. Allostatic drive operates to produce variation rather than homogenization. In the case of sensorimotor systems, it is the interaction between peripheral systems and central nervous system mechanisms that determine whether or not the system will remain

---

[6]   a cylindrical, wheeled robot often used in sensorimotor-oriented robotics research.
[7]   the ability of a system to retain stability in the face of perturbation. In the literature, this also refers to the ability of a genomic or morphological system to maintain its function in the face of change. This change can be produced by evolutionary processes, or imposed by perturbation.
[8]   under conditions of perturbation and learning, "hidden" or previously latent variation is expressed via reactive or activity-dependent plasticity. This can result in a capacitance mechanism that initiates allostatic drive and changes the internal state accordingly. See Table 5 for a similar discussion.

in homeostasis or undergo allostatic drive.

Bundle et.al (2006) found that in sprint cycling, compensatory mechanisms exist for dealing with fatigue. In the case of endurance activities, metabolic processes such as tissue adaptation may play a complementary role to neuronal processes such as changes in the recruitment of motor units and learning and memory. Functional MRI (fMRI) research focusing on these neuronal processes provides a clearer picture of how these compensatory processes operate. Liu et.al (2002) found that continuous force production can elicit muscle fatigue, which has similar effects on force output and measurements of muscle activity, but involves significant modulation by cortical brain regions.

*2.1.4 Definition of Mechanisms of Physiological Regulation.* A more direct connection between behavior, physiological regulation, and the functional phenotype can be demonstrated by perturbing specific mechanisms that constitute the internal model (Kawato, 1999). For example, non-associated stimuli that block the effects of perturbation will prevent the internal model components from stabilizing behavioral output. The effect of such stimuli can be simulated using pharmacological agents and lesions. For example, the GABA-A agonist muscimol[9] (Krakauer and Shadmehr, 2006) can act to shut down sensorimotor learning acquisition and consolidation in the cerebellum, while NMDA[10] antagonists have been used to inhibit sensorimotor learning consolidation in the parietal and sensorimotor cortices (Boniface and Ziemann, 2003). In cortical regions, the sprouting of new synapses[11] has been found to occur as a correlate of behavioral improvement during training (Selzer et.al, 2006, Pages 238-240; Mellor, 2006; Sweatt, 2003, Page 95). When initiated and sustained by technological interventions, these changes can affect the stability of homeostasis.

Lesion experiments have shown that functional activity within certain portions of the proposed internal model can have a direct effect on consolidation and recall (Ishibashi et.al, 2002). Specific to the context of tool-use and performance, localized activation in the cerebellum and posterior parietal cortex (Shadmehr and Brashers-Krug, 1997) can be associated with training using computer mice of variable shapes and levels of tracking gain. An RT-PCR[12] analysis of posterior parietal cortex in a rhesus monkey revealed the upregulation of gene products and receptors during but not following training that involved reaching for food with a rake. After the surgery, subsequent learning of targeted reaching behavior using the same arm was severely impaired. These results were due to the inability to encode proprioceptive feedback from the hand and arm, which was impaired after these tissues and associated functional attributes were removed (Ishibashi et.al, 2002).

Alternating between weighted and unweighted contexts can also have an effect on how the internal model and associated nervous system activity produces an output signal to the limbs. In recordings and stimulation of the motor cortex in non-human Primates, Graziano et.al (2005, Page 108) show that a 90g lead bracelet has an effect on how far the individual can lift their arm upward. The difference in range of motion between a free lift and lifting with a weight on the arm is interpreted by the authors (in Graziano, 2009, Page 108-109) to represent a lack of compensatory mechanisms. With electrical stimulation, the individual can raise their arm to the same height as occurred before weighting.

---

[9] a selective agonist (a molecule that binds to a specific receptor type to elicit an electrochemical response in a neuron) for GABA-A receptors. This neurochemical, a psychoactive drug derived from mushrooms, binds to GABA-A receptorsand has the effect of "turning off" the part of the brain to which it is applied. Used in eyeblink conditioning experiments, it can elicit effects on the cerebellum during memory acquisition.

[10] N-Methyl d-Aspartic Acid, a form of Aspartate. NMDA acts to facilitate neuronal communication at the synapse. NMDA is a ligand (a substance able to bind to a specific molecule for a specific biological purpose) that activates a specific subtype of glutamate receptor.

[11] a form of neuroplasticity that operates in response to leaning. Synapses are gapped connections between neurons that facilitate the physical transfer of information between them. The growth of new synapses is mediated by receptor-dependent plasticity and novel information that needs to be encoded.

[12] Reverse transcriptase polymerase chain reaction. This is a means to measure the synthesis and abundance of mRNAs in a particular tissue. The mRNAs are reverse-transcribed into nucleotide fragments, which are further sequenced and replicated using regular PCR. Demonstrates elevated expression (upregulation) of specific gene products.

*2.1.5 Definition of sensorimotor learning, memory, and integration.* Multisensory phenomenon that requires iterative sensorimotor integration are an integral part of the internal model (Shadmehr and Wise, 2005, Page 2; Oie et.al 2002). For example, the initiation of learning involves mechanotransduction, or the integrating information from mechanical and touch receptors in the joints and limbs with a supervisory role for visual perception (Wolfe et.al, 2005, Chapter 12). Sensorimotor learning, like forms of declarative memory, exhibits distinct stages of acquisition, consolidation, and recall that can be altered by the introduction of novel environmental conditions (Shadmehr and Holcomb, 1997; Donchin and Shadmehr, 2004). In this paper, the internal model that drives these behaviors is considered from a systems-level point of view, which is relevant to many domains of applied research. This internal model is characterized by a series of brain regions and physiological mechanisms that includes centers involved in the major functional attributes of adaptive sensorimotor behavior (Graziano, 2009, Chapter 7).

*2.1.6 Contributions of skeletal muscle adaptation and standing phenotypic variation.* Skeletal muscle adaptation and standing variation [13] in limb morphology can also lead to the achievement of sensorimotor ergonomy and perhaps even augment sensorimotor learning. Payton (1974) found that muscles in the dominant hand and arm responded to sensorimotor learning during shuffleboard playing. Moreover, Aoki (1990) found that muscles responsible for flexion in the dominant arm also respond to rapid angular movements at the wrist. Finally, a greater number and volume of mitochondria [14] and an associated reduction in the accumulation of muscle lactate, which affects muscle activity and other metabolic factors across training (Wu et.al, 1999). These changes may affect electromyographic (EMG) recordings.

In terms of standing variation, there is a role for morphological dampening (Full and Koditschek, 1999; Nishikawa et.al, 2007). Morphological dampening can be defined as the ability of a morphological structure to assist in the manipulation of a tool or wearable device, and occurs when body morphology enforces a physical constraint upon the output of physiological systems. One specific mechanism that acts as a physiological limit to human performance is the allometric scaling of phenotypic traits [16] (Cheverud, 1982). In adult organisms, scaling of the limbs generally minimizes the energetic requirements for movement (Full and Koditschek, 1999). For example, the ratio of shank length to thigh length can be optimized at a value of 1.06 in stripped-down physical models of gait such as passive dynamic walker robots (Collins et.al, 2005). This is consistent with what is seen in humans with a normal gait. Various scaling relationships may contribute to a differential capacity for adapting to environmental perturbations, as those perturbations can either be dampened at the level of limb

---

[13] variation in the phenotype that is the product of evolution (e.g. selection and neutral processes), but is observed only in the current generation.

[14] cytoplasmic elements with their own circular genome that play a key role in synthesizing ATPs, which supply energy to the cell. When there is an increase in mitochondria production, there is a corresponding increase in the production of ATPs, which provides a surplus of energy for metabolic processes.

[16] allometry can be defined as the relationship between two phenotypic measurements, which can also be thought of as a proportion. These proportions are variable across individuals within a species, between species, and development due to genetic variation and linkage.

configuration, or drive adaptation in skeletal muscle and neural systems.

**2.2   Theoretical Model of Performance Augmentation.** The theoretical model introduced in this paper brings together behavioral phenomena with its anatomical and cellular-level correlates as they relate to technological interaction. Traditionally, the idea of an "internal model" involves the integration of physiological behavior with cognitive and physical information, the dimensions of which have been shown to be stored in certain cell populations (Koch, 1999, Chapter 18; Aflalo and Graziano, 2006). One way to do this is to propose an internal model divided into three modules within which the learning, memory, and integration of environmental and physiological state information occur.

The first component of this model involves the prefrontal cortex, visual system, striatum, and sensorimotor cortices, and plays a supervisory role for the planning, representation, and integration of proprioceptive information (Georgopoluos *et.al*, 1982; Porro et.al, 1996; Roland et.al, 1980). This is an important factor in the design of teleoperation[17] systems, where Sheridan (1993) recommends that supervisory control and impedance control of the arm are critical for maintaining consistent performance in human-machine systems. These parts of the brain are also the source of prism adaptations, which are temporary modifications to the visual system that may initiate sensorimotor forms of learning (Fernandez-Ruiz and Diaz, 1999).

The second component involves the medial parietal cortex and the cerebellum. This relates to the effect learning a new set of environmental conditions has on adaptation and plasticity. An alternate means of producing this manipulation is to apply rotational forces to a specific anatomical segment while the movement is being executed. Several researchers (Conditt et.al, 1997; Shadmehr and Mussa-Ivaldi, 1994) have applied Coriolis forces to the arm while movements to a target location were made using a joystick. When the force field is changed but all other elements of the experimental setting remain the same, the operator must relearn how to activate their muscles in the context of this change. This has been shown by experiments using a centrifugal drum to simulate weightlessness (Lackner and DiZio, 2000), and experiments where humans throw clay balls at a target while on a merry-go-round (Martin et.al, 1996).

The third component involves the thalamus, the spinal cord, and relevant muscles in the peripheral nervous system. These components produce force output based on both local and central nervous system information. One example of this are preflexes (Brown et.al, 1995), which produce a local closed-loop response in the spinal cord. Afferent (e.g. feed-forward) information in the spinal cord travels to the nervous system via two routes: the DC-ML (dorsocolumnar-medial lemniscus) pathway, and the spinocerebellar pathway. The former ultimately delivers touch and vibratory information to cortical areas, which are involved in processes such as actively maintaining balance (Jeka, 2006) and reaching and pointing (Shadmehr and Wise, 2005, Pages 70-74. Afferents from this pathway project to the thalamus in the contralateral hemisphere (e.g. the side of the body opposite to where movement occurred), and carry information utilized in determining conscious states and cognition. The latter pathway delivers tension and muscle contraction information to the ipsilateral cerebellum (e.g. afferents in this pathway do not cross the midline of the body). This information is involved with the unconscious aspects of movement (Shadmehr and Wise, 2005, Pages 70-74). In both cases, proprioceptive inputs are important because they convey positional information, the activity of joints, muscles, and a given limb end-effector's spatial position (Morasso and Sanguinetti, 1994; Dietz

---

[17] a means of controlling a computer or mechanical system remotely. One example of this type of system involves controlling a virtual environment with a remote device. Teleoperation usually involves a mapping of commands and instructions from the human body to a technological device.

et.al, 2004). One example of this interaction is the production of motor primitives[18] (see Box 4), which are assembled at the level of the spinal cord to modulate complex movements (Mussa-Ivaldi, 1999).

**Box 4. Definition of motor primitives and preflexes**

In the robotics and neuromechanics literature, there are two concepts which emphasize how complex movements can be produced by relatively simple mechanisms. In Giszter et.al (1993), all limb movements in frogs can be reduced to a set of 12 non-adaptive primitives. These primitives are similar to the central pattern generators (CPGs) responsible for insect flight (Delcomyn, 1980). Complex and highly-tuned movement behaviors can arise by recombining these primitives in various ways. Hamburger et.al (2006) modeled this set of motor primitives using a robotic system to produce a standing-up behavior. To attain stable standing-up behavior, the system had to both superimpose and sequentially execute the primitives into a single, complex movement.

In the case of vertebrates, these primitives are thought to be assembled in the spinal cord, and then used to produce feed-forward movement (Giszter et.al, 1993). Depending on the criterion and methods used, motor primitives may originate in the olivocerebellum rather than the spinal cord. In his seminal theory, Marr (1969) proposes that memory and sensorimotor integration mechanisms governed by the output of Purkinje cells is in part based on climbing fiber inputs from olivary cells. Olivary cells may themselves exhibit an encoding for a single component of a complex movement, which Marr refers to as 'pieces of output' or elementary movements (Marr 1969). A more recent concept called a "preflex" (Brown et.al, 1995) suggests that combinations of motor primitives can be encoded in the central nervous system as a single functional unit. This information, which operates in identical fashion to motor reflexes, can override the production of single motor primitives as the fundamental basis for feed-forward limb movement (Prochazka et.al, 2007; McKinstry et.al, 2006). Whether a preflex is active or not depends upon the effects of learning and memory and inertial feedback.

Aside from information pathways in the spinal cord, the third module is affected by mechanotransduction, muscle plasticity, and limb morphology, and can result in the encoding of preflexes[19]. All of these phenomena occur in the periphery, and influence processes that occur in the first two modules. Feedback is accomplished when information along this efferent pathway is returned to the brain network. Such internal feedback has been referred to as an efference copy (von Holst and Mittelstaedt, 1980, Pages 176-209; Bridgeman, 2007), and serves as a means for the brain network to compare the state of action at subsequent points in time and adapt its output accordingly.

*2.2.1  Perturbation.* The introduction of a perturbation is done during either learning block 1 or 2, and results in two outcomes: triggering or interfering with processes associated with memory consolidation and the progression of physiological adaptation. Perturbation can be mathematically defined as

$$P = d_{s1} - d_{s0} / t \qquad [1]$$

where P is the degree of perturbation, $d_s$ is the magnitude of perturbation related to inertial feedback at a particular point in time, and t is time characterized experimentally by the number of trials during which perturbation is presented. This is similar to a traditional rate equation.

---

[18] a concept that explains the regulation of complex movements in vertebrates, particularly in tetrapods. According to this model, complex movements arise from a series of brief muscle twitches that produce a certain amount of force and directional movement. Information about these primitives are encoded in the central nervous system and spinal cord.
[19] a complex, zero-delay movement (Holmes et.al, 2006) that is produced in an automatic fashion (e.g. like a reflex). According to current theory, preflexes are encoded in the cerebellum and other regions of the brain (McKinstry et.al, 2006). However, preflexes are also influenced by limb morphology and inertial feedback.

**2.3 Systematically Perturbing the Environment: cues from environmental medicine.** Applying methodology from the field of aerospace and environmental medicine has the potential to answer questions regarding physiological adaptation to hybrid bionic systems in simulations of physical environments and the role of learning, memory, and mechanotransduction in this process. Traditionally, the aerospace and environmental medicine approaches involve perturbing the environment and investigating subjects with a control physiology. One example of this is the unloading paradigm, which simulates the effects of weightlessness on muscle, bone, and force production by introducing a distorted set of environmental forces during the presentation of a perturbation (Mozdziak et.al, 2001). Such results have been applied to areas as diverse as deep sea exploration, space travel, and jet fighter aircraft design. This differs from regular medical studies in which a normal environment is used to probe abnormal physiologies. Table 1 shows the different ways in which physiology and environment might be manipulated in a combination of such approaches. This approach can be modified in a number of ways, including probing the relationship between physiological homeostasis and adaptation to cold water submersion (Wissler, 2003) and high-altitude hypoxia training (Rice et.al, 2003).

**Table 1. Perturbing physiology and environment using the aerospace medicine paradigm.**

|  | **Normal Physiology** | **"Mutant" Physiology** |
|---|---|---|
| **Normal Environment** | Nothing perturbed | Maintain environmental information, perturb internal processes |
| **Abnormal Environment** | Perturb environmental information, maintain internal processes | Perturb environmental information, perturb internal processes |

In Table 1, a distinction is made between normal and abnormal environment. These distinctions were made to illustrate the environmental medicine approach. Likewise, there is a distinction between normal and "mutant" physiology. Mutant physiologies might include genetic mutants, physiological systems ablated in development, or represent different polymorphisms tied to variation observed in behavioral and physiological indicators or plasticity in those indicators. A range of normal and mutant physiologies can be tested using the environmental medicine paradigm, which would be the case when accounting for individual variation in performance. The combination of sensorimotor learning and environmental physiology techniques allow for the structural, phasic changes in behavior and physiology that accompany performance augmentation to be observed.

One way to quickly and effectively expose individuals to an abnormal environment using virtual reality is to introduce a novel activity for a period of time, and then interfere with learning using a systematic perturbation. To uncover the robustness (see Box 5) of a particular physiological state, an experimental setting that introduces differential amounts of proprioceptive information must unveil two features of the physiological system: how quickly the parameters of a perturbed environment can be acquired and consolidated, and how easily intermittent exposure can disrupt a memory trace for a physical behavior performed in a non-perturbed environment.

**Box 5. Definition of robustness and internal state.**

Morasso et.al (2005) have identified the preflex as a fundamental unit of adaptation in sensorimotor physiology. Furthermore, McKinstry has identified preflexes as internal states shaped by the constraints of the phenotype and inertial feedback. In Box 4, the results of Hamburger et.al (2006) were introduced which replicated the recombination of motor primitives to achieve stable standing-up behavior. Their results revealed that some combinations of motor primitives produces standing-up behavior that are prone to failure (e.g. falling down), while other standing-up behaviors produced in this way were much less prone to failure over repeated trials. Phrased in terms of robustness, different combinations of motor primitives (or preflexes) provide different levels of resilience to perturbation. Some configurations will not provide adequate levels of control and stability in the face of perturbation. The movement of the internal model from one state to another is contingent upon the robustness of the original internal model. To meet the needs of both sets of conditions, the internal state selects a series of motor primitives, encoded as a single preflex, that best meets the need of both sets of environmental conditions. In order to accommodate these new conditions, the internal state must exhibit allostatic drive, which in this case translates into adapting to a series of new, more inclusive states.

**2.4 Prosthetic Interface Design: way to create a simple hybrid bionic system.** What is required is a heuristic means to understand the proposed internal model as it relates to the achievement of sensorimotor ergonomy. To do this, two prototypes of such a tool were created (see Figure 1 and Methods section). The purpose of this tool is to simulate a set of environmental conditions that perturb upper-arm morphology and the current physiological state.

### 3.0 Research Questions and Hypotheses
**3.1 Aspects of the Sensorimotor Environment.** In applying sensorimotor learning and brain networks function to a specific set of conditions, there is interplay between physiological activity and computing environments. As models of phenomena encountered in the physical world, computing environments introduce novel sets of physical conditions that must be adapted to by the operator while also being able to return to alternate physical conditions. One such example involves pilot or driver training, as operators often train on a computer simulation, perform the activity repeatedly on the physical interface, and then return to the simulator for additional training. There are physiological consequences of this long-term system use that requires a plastic response from internal processes and mechanisms to produce either a fixed or bistable behavioral response.

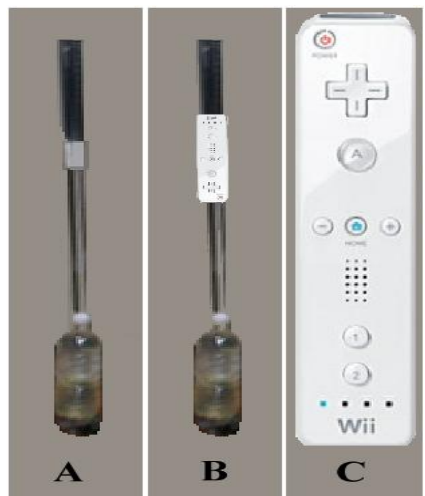

Figure 1. Picture of prosthetic interface designed to alter environmental forces encountered during behavior. From left: prosthetic device alone (A, not used in experiment), loaded controller (B, motion controller attached to prosthetic device), and naked controller (C).

**3.2 Main research questions.** There are two basic questions that drive the experiment featured here. Each question serves as motivation for the rest of the empirical section.

**Q1:** How does perturbing environmental information contribute to a phenomenon called allostatic drive, and how does allostatic drive contribute to the overall coupling of this system?

**Q2:** How does introducing perturbations at different points in the experimental sequence effect homeostasis and allostatic drive?

A hybrid bionic interface is a system that interacts with or otherwise affects the nervous system. This experiment, by being both immersive and taking into account physiological parameters, serves as a surrogate for what hybrid interfaces might become as they move from laboratory to the marketplace.

**3.3 Reason for answering these questions.** This experiment and approach in general addresses how physiological systems bounded by the phenotype and technological systems adapt to environmental perturbations. Augmentation involves interactions between internal processes and environmental challenges. Of particular importance is behavioral, physiological, and morphometric conceptual results and data in determining if and how technological system become isomorphic to the internal physiology.

**3.4 Hypothesis A: Augmentation by Subsequent Effects.** The **Augmentation by Subsequent Effects Hypothesis** states that different types of technological environments will act to augment human performance. As hybrid bionic systems become more prevalent, the need for a seamless performance transition from hybrid to non-hybrid environments and between different kinds of hybrid environments will become critical. Operational examples of such systems include helicopter cockpits and deep-sea exploration vessels. Perturbation will have an unmeasured effect on homeostasis, which will in turn have an effect on performance level.

$H_A$: Switching induced by alternating technological environments has an effect on performance. This effect unfolds across multiple blocks of experimental trials.

This hypothesis deals with the degree of inertial feedback provided by the loaded controller versus the relative lack of such provided by the naked controller. Explicitly, inertial feedback provided by the loaded controller provides a greater amount of information content, and thus greater information transfer to the nervous system and brain network.

**3.5 Hypothesis B: Environmental Information.** The **Environmental Information hypothesis** states that perturbation will have an effect on homeostasis and performance level.

$H_B$: Environmental perturbation with the prosthetic device results in increased variability of performance level measurements such as muscle activity and mapped physiological output, and contributes to allostatic drive.

The manner in which the environment responds to direct manipulation has a direct impact on available sensorimotor environmental information. The homeostasis measurement was modified slightly to measure the interval between two baseline measurements. In the parlance of control theory, this type of perturbation either establishes a set-point or changes it so as to induce allostatic drive. The expected difference in performance between homogeneous and perturbation conditions is similar to an affordance, which has been defined by Gibson (1979, Pages 133-143) as behavior constrained by the information content of a specific environmental context. In the case of performance augmentation, the less environmental information available, the less variability observed in the indicators of muscle activity, which maintains the current homeostatic state.

**3.6 Hypothesis C: Functional Static Allometry.** The **Functional Static Allometry hypothesis** states that the bounds of adaptability will have an effect on performance level.

$H_C$: The bounds of adaptability as characterized by the forearm-humerus ratio used as a predictor of performance level will produce a nonlinear scaling function for each experimental condition.

The two phenotypic assays in question are body height and forearm length. The ultimate causal mechanisms that account for these differences are based on the interplay between morphology and the individual ability to adapt to current environmental conditions. The functional morphology of upper-body limbs is expected to yield an optimal state among specific scaling ratios as has been demonstrated in the design of biologically-inspired robots (Hara and Pfeifer, 2003, Page 59). Such differences are also expected to yield optimal performance for certain scaling values.

**3.7 Hypothesis D: Allostatic Drive Hypothesis.** The **Allostatic Drive hypothesis** states that perturbation will have an effect on the homeostasis measure.

$H_D$: The combination of inertial feedback and perturbation will have a negative effect on the maintenance of homeostasis based on the performance level measures.

Based on what is known about neuromuscular systems, increased skeletal muscle activity in the form of perturbation and representation of a stimulus leads to adaptive changes at both the molecular level and in the way in which motor units are innervated and recruited by motorneurons. Thus, the muscle activity measurement should increase in the loaded controller homogeneous and loaded controller perturbation conditions, and decrease with the retention of learning.

## 4.0 Methods

**4.1 Experimental Design.** An experiment was conducted to probe the augmentation and adaptability of human sensorimotor performance and physiology. The experimental design was a cross between the motor learning (Mussa-Ivaldi, 1999) and aerospace medicine (Rafiq et.al, 2006) experimental paradigms. A 2 (prosthetic device) x 2 (perturbation type) x 3 (experimental blocks) x 16 (trials) mixed experimental design was used. The between subjects factors are prosthetic device and perturbation, while the within subjects factors are learning blocks and trials. Prosthetic device setting has two levels: naked controller and physical controller. Perturbation type has two levels: homogeneous and perturbation. Learning type has three levels, learning block 1, learning block 2, and learning block 3. Table 2 shows the specifics of this design.

**Table 2. Operator Dynamics: experimental design layout of the 2x2x3x16 design\*.**

|  | Prosthetic device (Naked) | | Prosthetic device (Loaded) | |
|---|---|---|---|---|
|  | **Naked Controller Perturbation During Learning Block 1** | **Naked Controller Perturbation During Learning Block 2** | **Loaded Controller Perturbation During Learning Block 1** | **Loaded Controller Perturbation During Learning Block 2** |
| Block 1 | Naked | Loaded | Loaded | Naked |
| Block 2 | Loaded | Naked | Naked | Loaded |
| Block 3 | Loaded | Loaded | Naked | Naked |

* 16 trials were administered for every block. This is equivalent to a single block.

## 4.2  Participants.

Thirty-seven participants were used in this experiment. Participants gave their informed consent before participating in the study. The study was approved by the local ethics committee and performed in accordance with local ethics committee standards.

## 4.3  Apparatus.

The simulation portion of this experiment was conducted using the Nintendo® Wii gaming platform (Nintendo Corporation, Kyoto, Japan) using the simulation *Wii Sports Golf* (see Figure 2). The Nintendo® Wii uses several simulation simulation-specific parameters and a motion controller to produce computer-generated simulation action. All simulation-specific parameters, such as wind speed and club surface type, were held constant.

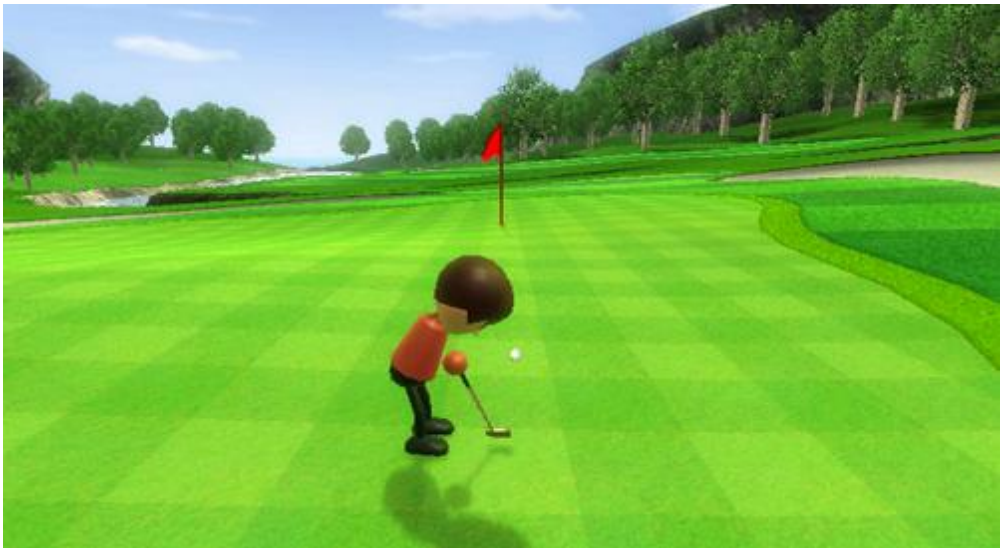

Figure 2. Screen shot of Wii sports golf simulation.

*4.3.1  Prosthetic Manipulation.* The prosthetic manipulation was achieved by varying the type of input device use for the motion controller. The Wii motion controller uses a gyroscope to translate actions produced by the human into a virtual analogue of physical movement. While this produces inertial gyroscopic forces of limited scope which act as a source of inertial feedback, the gravity and velocity effects of the simulation physics should be consistent within each condition. The experimental manipulation of the environmental variable was achieved by the use of two types of controllers.

*4.3.2  Naked controller conditions:* The naked controller conditions will involve using the controller that is standard with the Wii system. That is, the operator will manually use the motion controller, mimicking the motions of the reaching activity without any of the attendant inertial feedback from the coupled device (i.e. motion coupling between the prosthetic device and counterweight).

*4.3.3  Loaded controller conditions.* The loaded controller was used to operationalize environmental perturbation. The loaded controlled consists of a Wiimote motion controller strapped to a customized prosthetic device. In this experiment, the prosthetic device consists of two golf clubs bound together at the shaft with a dynamic counterweight (a bottle filled with a liquid) attached at the base of the shafts. This dynamic counterweight is referred to as a forcing chamber. The loaded controller prosthetic device was designed this way to ensure that previous experience with golf clubs or other swinging devices was minimized and that an environmental perturbation could be systematically introduced.

The loaded controller provides a significant perturbation to an activity performed in a virtual environment, and provides a distortion of proprioceptive information that complements the absence of this type of information presented by the naked controller. During both the naked and loaded blocks, the operator will make sixteen reaches with the prosthetic device. In the simulation, the action will take place on a putting green. The goal of each trial was to get the ball in the hole as often as possible. When successful, this will minimize the variance of the mapped physiological output measurement for a single trial.

*4.3.3 Phenotypic state at the end of development.* Phenotypic assays can be used to determine how physical performance and augmentation are limited by growth regulation that occurs during development. Changes observed in the sensorimotor regulatory system are in part constrained by the overall size and shape of the forelimbs. As the physics of the forelimbs at the end of development interact with the variable physics of the environment, they either act to preserve the internal state or contribute to homeostasis.

**4.4 Measures**

*4.4.1 Physiological Activity.* To collect electrical potentials from the nervous system and information about internal physiological states, the Biopac MP150 biopotential amplification system was used (Biopac Systems, Goleta, CA). Two channels was used to collect these data from the muscles of the forearm and humerus.

A template was created in AcqKnowledge® (Biopac Systems, Goleta, CA) to digitize, smooth, and transform the raw data in real time. A sampling rate of 1000Hz was used to collect the raw data. A calculation channel with a window of 10 data points was used to smooth the muscle activity data using a moving average routine which resulted in a final sampling rate of 100 data points per second. A special calculation channel was used to apply an integrated EMG (iEMG) filter to rectify and mathematically integrate the signal representing muscle activity. The hardware filters were set to a gain of 5000, a high-pass analog filter value of 1Hz, and provided notch filtering of the signal (50dB rejection mode at 50-60Hz). A notch filter was used to remove ambient spectral noise.

*4.4.2 Muscle activity.* Electromyography (EMG)[20] was collected using the Biopac MP150 system (Biopac Systems, Goleta, CA) to quantify muscle activity. Multiple segments act semi-independently to determine the consequences of sensory input, feedback, and variability. Two points on the upper arm were selected to record EMG data: the flexor carpi radialis[21] along the forearm and the triceps brachii[22] along the humerus. Muscle activity at these locations can be used to determine arm movements in relation to dynamic upper-body posture (see Figure 3).

The muscles recorded in these experiments were selected based on two factors: muscles that were located along the dorsal or ventral surface of the arm, and muscles that were involved in the complex reaching movement. Several steps were taken to ensure the internal validity of these measurements[23].

---

[20] Electromyography (EMG) is a method of recording electrical activity from specific muscles to quantify changes in firing relative to anatomical movement. The power of an EMG signal is roughly equivalent to action potentials fired in response to reflexive behavior or anticipatory single sent from the sensorimotor brain centers.

[21] A muscle of the forearm that acts to flex (contract) and abduct the hand. Flexor carpi radialis inserts proximally at the medial epicondyle of the humerus and distally at the base of the fifth metacarpal.

[22] A muscle of the humerus involved in extension. Triceps brachii helps to keep the arm straightened in the face of resistance, and inserts distally at the olecranon process of the ulna.

*4.4.3 Mapped Physiological Output Measurement.* A measurement of mapped physiological output was calculated for several reaches over the course of a two-minute trial. This is a measure original to this document. While it is not a direct measure of force, it is related to how afferent signals from the nervous system are attenuated while virtual objects are being controlled in a virtual environment. Mapped physiological output was measured using the following equation

$$\mathbf{MPO}_t = |\mathbf{D_{req}} - \mathbf{D_{moved}}| / \mathbf{D_{req}} \qquad [2]$$

where $D_{req}$ is the fixed distance a virtual object needs to be moved over a given trial, $D_{moved}$ is the distance the virtual object is actually moved resulting from muscle force production captured by the input device, and $MPO_t$ is the mapped physiological output for a single reach using the prosthetic device. A single reach will consist of taking a swing with the prosthetic interface. The operator will start each swing aligned with markers on the floor that denote a baseline or resting physical state. The distance the ball travels in the simulation and a completion time constant was collected. For each reach, the simulation was reset so that the shot is always taken from the beginning of the first hole. Sixteen reaches were administered across a single experimental block.

*4.4.4 Phenotypic assays.* Phenotypic assays were collected for each participant prior to the experimental session (see Figure 3). The following measurements were taken: forearm, humerus, lower body height, and total height. All measurements are taken prior to being seated at the interface. A measuring tape was used to collect all measurements listed herein. The measurements listed in this section are the measurements that were directly measured from the participant. The following measurements were calculated from the empirical data: total arm length (AL), humerus length (H), head and trunk (HAT) height, shank length (SL), and thigh length (TL).

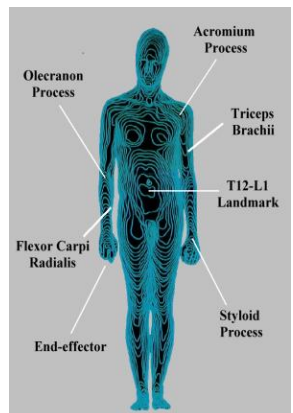

Figure 3. Configuration of the major anatomical landmarks, muscles, and interaction points explored in this document.

---

[23] The muscles were located using a digital musculoskeletal atlas (http://rad.washington.edu/atlas), measurements of the arm's surface, and palpatation. These locations were confirmed by testing the preparation before recording. The trace recording was assessed for muscle spindle activity and ECG artifact during one instance of complex reaching. If any abnormalities were detected, the preparation was redone along with relocation of the surface electrodes. The connections were checked with an impedance meter before the experiment began and periodically in between blocks.

The total height measurements involve subjects standing in an upright posture with their backs against the wall. A yardstick was affixed to the wall immediately behind the subject. The total height (TH) measurement was taken from the feet to the saggital crest at the top of the head. This measurement was measured before the experiment and used as a constant in calculating the HAT measurement. For measurements of the arms and upper body, subjects will remain standing but away from the wall. The method of Martin and Nguyen (2004) was used to assay arm length and biacromial breadth, head and trunk (see Winter, 1990, Chapter 4), and lower body measurements. Participants will extend their arms, hands, and fingers maximally outward from their sides so that the subject's wingspan can be measured.

Using this pose, the experimenter obtained four upper-body measurements. The first is shoulder breadth, defined as the distance from the left acromioclavicular (AC) joint to the right AC joint. The second is total wingspan, defined here as the distance from the left styloid process (distal end of the ulna) to the right styloid process (distal end of the ulna). The third measurement is forearm length, which was taken on both the right and left arms from the elbow joint (olecranon process - proximal end of forearm segment) to the styloid process (distal end of the ulna – distal end). The fourth measurement is humerus length, which was taken on both the right and left arms from the olecranon process (elbow joint) to the AC joint.

Measurements should be straight-line, one-dimensional assays along the dorsal surface of the coronal plane. The total arm and humerus lengths can be derived from these measurements. Total arm length (AL) can be calculated by the following equation

$$\textbf{AL = (TWS – BB) / 2} \qquad\qquad \textbf{[3]}$$

where TWS is the total wingspan and BB is the bicromial breadth. Humerus length (H) can be calculated as

$$\textbf{H = AL – F} \qquad\qquad \textbf{[4]}$$

where AL is the arm length and F is the forearm length.

## 4.5 Constructs and Conceptual Measurement

*4.5.1 Performance level.* Performance level serves as a dependent variable. A decrease in mean EMG spectral power and decreased behavioral variability during each block of an experimental condition as reflected in mapped physiological output can often be considered different dimensions of improved performance. Generally, performance level can be defined using the following equation

$$\textbf{L}_\textbf{p} = \overline{PM} \textbf{ / T} \qquad\qquad \textbf{[5]}$$

where $L_p$ is the level of performance, $\overline{PM}$ is the performance measurement (taken as an average over multiple trials), and T is the number of trials being averaged. The performance measurement represents either the mapped physiological output calculation or aggregate spectral transformations of the physiological recordings. These are complementary measures of performance improvement or degradation given the current environmental state or experimental phase. When performance level is compared across experimental phases, as in the case with decreased EMG activity, it is also a measure of adaptation.

*4.5.2 Inertial Feedback.* A concept related to unlearning is called inertial feedback. Mechanical sensory information provided by movement generated against an environmental medium can best be conceptualized as a volume of information that envelops the individual and acts as a "medium" enveloping the operator (see Snyder et.al, 2007 for application). Turbulence caused by force generated against this medium may play a key role in augmenting the function of sensorimotor integration mechanisms in the central nervous system (Nishikawa et.al, 2007). In the case of organisms that are entirely immersed in a specific environmental medium such as water, this information is quantified by the Reynolds number (Bejan and Marden, 2006), which approximates the viscosity and resistance of the body morphology to the environment as it acts against the body. To approximate the characteristics of this fluidity and resistance for the upper limbs in the context of mutating the environment, a forcing chamber on a prosthetic device was used. The inertial effects of the forcing chamber were kept constant by filling it with water at room temperature. This medium has a specific density of 1, which can be defined as

$$P \; \propto \; d_S, \; d_S \; = W \, / \, vol \qquad\qquad [6]$$

where P is the degree of perturbation, $d_S$ is specific density, W is the weight, and vol is the volume.

*4.5.3 Bounds of Adaptability.* The bounds of adaptability serve as an independent variable. The degradation of visual inputs in the experimental phases using the naked controller and limb lengths serve as boundaries to functional performance. One way the bounds of adaptability can be quantified is by mathematically scaling two or more phenotypic assays. One example of this is the scaling of total height to forearm length. When used as a predictor of performance level, it can be defined as

$$Y = \log(\text{-}Ax^2 + Bx - c) \q\qquad\qquad [7]$$

where Y describes a second-order polynomial. This particular definition of functional static allometry, which characterizes specific body sizes and shapes attained at the end of development, is based on growth regulation in development and may have an optimizing effect on performance (Garcia and Leal da Silva, 2006; Herr et.al, 2002). This variable serves as an aspect of the phenotype for specific experimental conditions.

*4.5.4 Homeostasis.* Homeostasis serves as a dependent variable. Operationally, homeostasis can be defined as the difference between EMG indicators taken either during or at the end of two subsequent experimental blocks. Mathematically, this can be expressed as

$$H = |BM_n - BM_{n+m}| \qquad\qquad [8]$$

where H is homeostasis, BM can be either the baseline or normalized block measurement (as long as the same indicator is used for both values), n is the first measurement point, and n+m is the second measurement point. Generally, when the absolute value of the difference between these measurements taken after learning block 1 and learning block 2 is minimized, the degree of homeostasis is high. When the absolute value of this difference is larger, the potential exists for allostatic drive. The maintenance of homeostasis across multiple measurement points is equivalent to robustness, while allostatic drive is equivalent to a compensatory capacity, or adaptability, to an experimental phase.

*4.5.5 Functional Fitness.* Functional fitness, or performance relative to the rest of the population, can be calculated as

$$F = |z_{min}| + z_{ind} / |z_{min}| + |z_{max}| \qquad [9]$$

where F is the functional fitness, $z_{min}$ is the z-score of the lowest performance level in the population, $z_{max}$ is the z-score of the highest performance level in the population (both based on a z-transform). In the case of MPO, $F^{-1}$ is used as the functional fitness. This results in a scale from 0 to 1, which provides a measure of population-level performance during an adaptive process (Nowak, 2006, Page 4).

*4.5.6    Cumulative Spikiness.* A measure of cumulative spikiness was used to quantify and detect performance variability in different portions of an experimental block. Spikiness is generally defined in adaptive physiological systems by the parameter *z* (Krishna et al, 2006), which is described mathematically as

$$z = max_i - min_i / mean_i \qquad [10]$$

where $max_i$ is the maximum value over interval i, $min_i$ is the minimum value over interval i, and $mean_i$ is the mean value over interval i. For the MPO measure, this construct was used to better detect patterns across the block by using applying these criterion recursively. Starting at the second trial, the interval i was equal to the *n*th trial in the block.

## 4.6    Procedure

The phenotypic assays were collected before the experimental session. The physiological measurements were collected using skin surface electrodes using an original procedure[24].

The participant was introduced to the experimental setting through a pretest training session. The operator wascome acclimated to the experimental setting using the naked controller configuration. This session will consist of eight trials. A two-minute long initial baseline will then be acquired, consisting of the participant swinging the dominant forearm like a pendulum and squeezing a styrofoam ball while flexing the dominant hand.

In the main experiment, the participant was randomly introduced to one of the between-subjects conditions. Each participant was randomly assigned to each condition. The participant experiences a different type of perturbation during either learning blocks 1 or 2, depending on what group the subject was assigned. For example, a participant might end up being assigned to the naked perturbation during learning block 1. For each condition, every block will be separated by a two-minute interval during which the resting baseline will be measured, measured in identical fashion to the initial baseline.

The naked perturbation during learning block 2 involves using the loaded controller for the learning block 1, a naked controller for learning block 2, and the loaded controller for learning block 3. By contrast, the naked perturbation during learning block 1 involved using the naked controller for the learning block 1, and a loaded controller for learning blocks 2 and 3. The loaded perturbation during learning block 2 involves using the naked, loaded, and naked controllers for the learning blocks 1, 2, and 3, respectively. Similar to the naked perturbation during learning block 1, the loaded perturbation during learning condition involves using the loaded controller for learning block 1, and then using the naked controller for learning blocks 2 and 3.

---

[24]  An impedance check was run using a Checktrode MK-III (UFI, Morro Bay, CA) unit on the preparation before and after each set of experimental blocks to ensure calibration of the instrument. The skin was cleaned and abraided using a mixture of 70% isopropyl alcohol (C3H8O), 30% water (H20), and electrode gel (Biopac Model GEL-1). Adhesive surface Ag-Ag-Cl electrodes (Biopac Model EL-503) were attached to appropriate places on the skin. The surface sites and portions of the electrode lead will then be secured in place with athletic tape; athletic tape was wrapped several times around the body segment in question. Both of these procedures was done minimize shifting of the electrode and lead wires and to maintain impedance between the skin surface and the electrode.

## 5.0 Results and Interpretation

**5.1 Statistical Tests and Analyses**. Basic statistical analyses were done in SPSS 14.0 (SPSS, Chicago, IL). Calculating descriptive statistics for the EMG signals and the sensitivity analysis simulations were done in MATLAB 7.0 (Mathworks, Natick, MA). Graphs and visualizations of the data were produced in Systat 11.0 and SigmaPlot 9.0 (Systat Inc., San Jose, CA). Basic statistical modeling involves conducting ANOVA analyses. Additional modeling techniques involved conducting regression analyses and curve fitting. The mapped physiological output measurement was averaged across all trials in an experimental phase. The analysis of EMG time-series data was done by extracting a segment representing all trials in each experimental phase and then running a signal processing transform on these data.

*5.1.1 Data Reduction.* The continuous physiological measurements were collected in raw intervals which were reduced using a moving average. The EMG data was further reduced by using an integrated filter. These data were further aggregated to an average over a single block of trials. Once all physiological signal data were aggregated, two transformations using signal processing methods was done: calculating the power spectrum using a Fast Fourier transform[25] (FFT – see Ramirez, 1984, Chapter 1) and finding the mean power frequency (MPF) for each integrated EMG channel. The mean power frequency is defined by Ritter et.al (2005) as

$$\mathbf{MPF} = \sum \mathbf{F} * \mathbf{P(f)} / \sum \mathbf{P(f)} \qquad \qquad [11]$$

where F is the range of frequencies over a particular interval of the signal, P(f) is a function representing the characteristic spectral power for each frequency, and $\sum$ is the integration of this range using the trapezoidal method. Naeije and Zorn (1982) and Gutierrez and Zarco (2003) have used mean power frequency for a particular interval of time to characterize aggregate the effects of perturbation on the neuromuscular system. This descriptive statistic will summarize nervous system activity so that comparisons can be made. For preprocessed EMG data, the FFT/mean power frequency approach is being used to characterize both continuous baseline activity and activity aggregated over several reaches during the eight trials of an experimental phase.

*5.1.2 Normalization of data.* In addition to providing information regarding changes in performance due to the previously presented block, the baseline measurements were also used to normalize all muscle activity measurements within each block. To do this, the mean power spectrum of the baseline measurement value was subtracted from the mean power spectrum value collected for each trial. This resulted in a certain percentage of correction in the signal that varied widely across individuals. Therefore, this correction made the muscle activity measurements more comparable across individuals.

**5.2 Basic Data Analyses: tests of significance.** An extensive series of analyses were carried out on these data. I conducted several repeated-measures ANOVA analyses on the mapped physiological output, transformed muscle activity, and transformed mapped physiological output data. To minimize the number of potential interactions, the data were placed into 12 groups for each measure. The 12 groups corresponded with each block of trials for all four conditions. This resulted in a 12 (block) x 16 (trial) repeated measures analysis for each measure. Additional repeated-measures analyses were conducted by analyzing the first two and last two condition separately. This resulted in a 6 (block) x 16 (trial) design, which yielded slightly different results than including all conditions in the same analysis.

---

[25] Before the Fast Fourier Transform, a low-pass (0-40 Hz) infinite impulse response (IIR) filter was passed over the integrated signal. This was done to reduce the amount of movement artifact. The Fast Fourier transform was conducted for each raw signal interval. The signals were zero padded, the magnitude of each frequency was calculated, and the filter was piecewise-linear. A Hanning window was used. These settings were used to reduce movement and ECG artifacts.

*5.2.1 Mapped Physiological Output (MPO) and muscle activity repeated-measures ANOVA.* For the 12 x 16 analysis, neither block (F = .747, p > .69) nor trial (F = 1.248, p > .23) were statistically significant. It was thought that one possible reason for this was the complex order of the perturbations and effects in the experimental design. To further minimize the effects of this complexity, the dataset was split in two. For a 6 x 16 analysis on conditions 1 and 2 only, block was nearly significant, F = 2.264, p > .054 (see Figure 4). For this same analysis, trial was not found to be significant, F = .966, p > .49. For a 6 x 16 analysis on conditions 3 and 4 only, block was not found to be significant, F = .358, p > .87. For this same analysis, trial was not found to be significant, F = 1.198, p > .12.

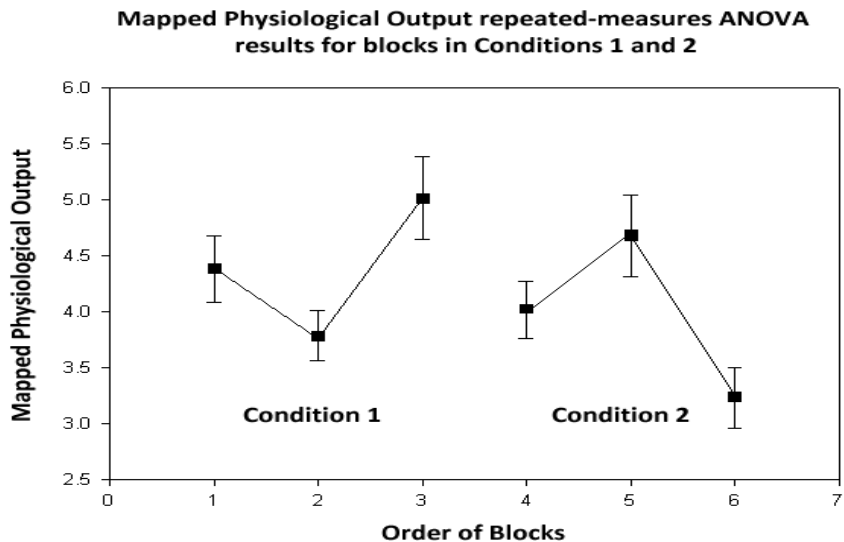

**Figure 4. Nearly significant results for blocks, 6 x 16 analysis of MPO measurement on conditions 1 and 2 only.**

*5.2.2 Muscle activity during experimental blocks.* A 2(condition) x 3(block) x 16(trial) repeated measures ANOVA was conducted for each muscle (triceps brachii and flexor carpi radialis) for condition 1 and 2 and conditions 3 and 4 separately. A total of four analyses yielded one significant relationship. In conditions 1 and 2, flexor carpi radialis (FCR) was significant, F = 4.082, p < .03 (see Figure 5). None of the potential interactions in any of these tests were found to be significant.

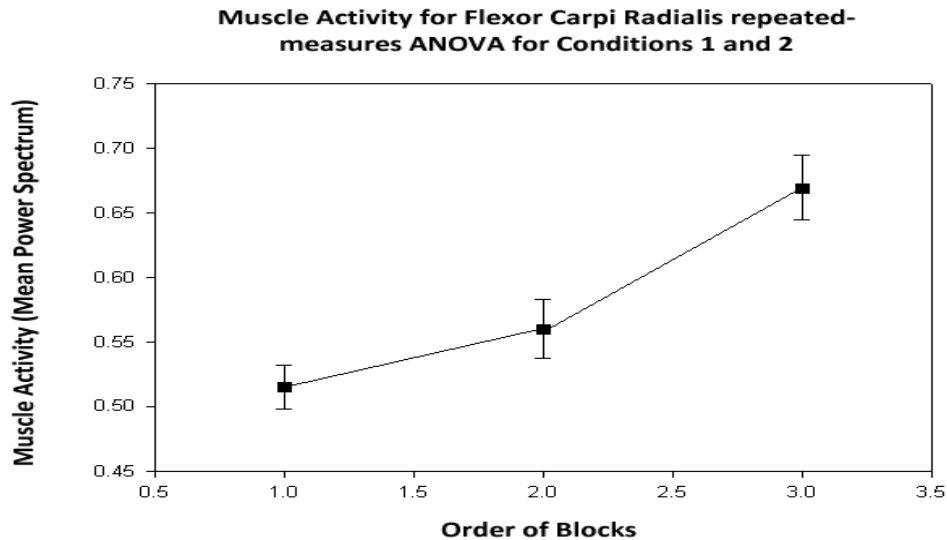

**Figure 5. Significant results for blocks, 2 x 3 x 16 analysis of FCR muscle activity for conditions 1 and 2 only.**

*5.2.3 Baselines and Homeostasis for muscle activity.* To test Hypothesis D, baseline and homeostasis measurements were analyzed using repeated-measures ANOVA. For the baseline and homeostasis measurements, two different repeated-measures designs were conducted: a 2 (muscle) x 2 (condition) x 3 (measurement) for the homeostasis measurement, and 2 (muscle) x 2 (condition) x 4 (measurement) for the raw baseline measurement. As with the MPO and muscle activity analyses, separate tests were conducted for conditions 1 and 2 and conditions 3 and 4. For the homeostasis measurement conditions 1 and 2 analysis, measurement was found to be significant, $F = 3.813$, $p < .035$ (see Figure 6). For the baseline measurement conditions 3 and 4 analysis, measurement was found to be significant, $F = 5.179$, $p < .005$ (see Figure 7).

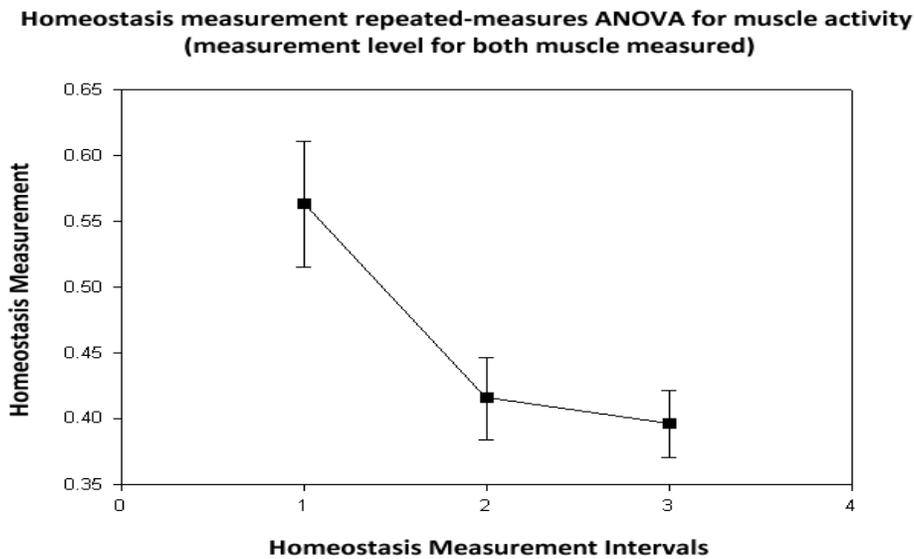

**Figure 6. Significant results for measurement, 2x2x3 analysis of homeostasis for conditions 1 and 2 only.**

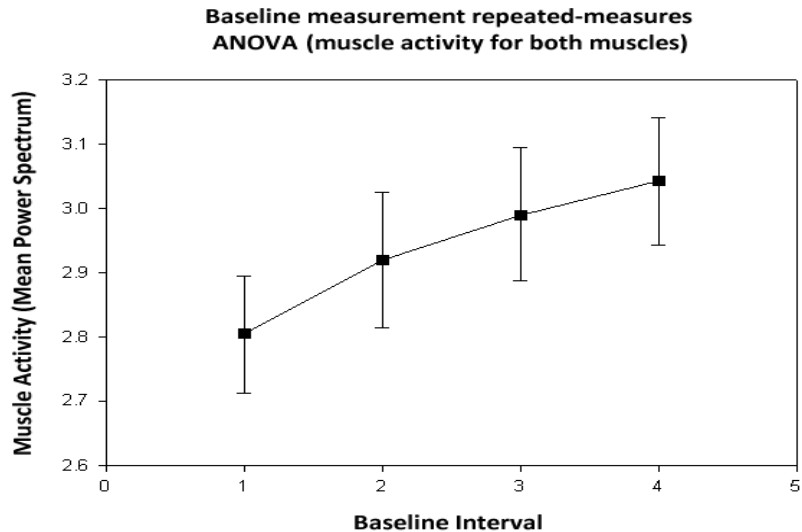

**Figure 7. Significant results for measurement, 2x2x4 analysis of baseline for conditions 3 and 4 only.**

*5.2.4 Mapped Physiological Output cumulative spikiness.* To assess how variable performance was across trials within individual blocks, the cumulative spikiness was calculated for each trial in every block (see Section 4.5.6)[26], and then analyzed using a 12 x 16 repeated-measures design used for the MPO analysis in section 5.2.1. Trials were found to be significant, F = 11.999, p < .001 (see Figure 8).

*5.2.5  Paired t-tests and correlational analyses for MPO and muscle activity.* To better understand what the pairwise relationships are between different experimental blocks, I conducted a series of paired t-tests (one-tailed, independent) and correlational analyses on each pair of blocks in each experimental condition. For the MPO and two muscle measurements, learning blocks 1 and 2, learning blocks 2 and 3, and learning blocks 1 and 3 were compared. In addition, the covariance for pairings of blocks for the MPO measurement was also calculated. The results of these analyses are shown in Table 3 and Figure 9.

**Table 3. Statistical Significances and Correlation Coefficients for each pair of blocks (MPO = mapped physiological output, TB = triceps brachii activity, FCR = flexor carpi radialis activity). All p-values < .05 level, correlation coefficients > .40 and < -.40, and covariances over 2.0 highlighted in red. * Paired t-test, one-tailed.**

|  | n = 19 | n = 16 | n = 16 | n = 19 | n = 16 | n = 16 | n = 19 |
|---|---|---|---|---|---|---|---|
| **Condition 1** | **MPO p-value\*** | **TB p-value\*** | **FCR p-value\*** | **MPO correlation** | **TB correlation** | **FCR correlation** | **MPO covariance** |
| Blocks 1 and 2 | .54 | .01 | .001 | .15 | .45 | .53 | 3.7 |
| Blocks 2 and 3 | .003 | .07 | .001 | -.01 | .43 | .52 | -.22 |
| Blocks 1 and 3 | .71 | .001 | .001 | .16 | .19 | -.06 | 5.11 |
| **Condition 2** |  |  |  |  |  |  |  |
| Blocks 1 and 2 | .06 | .001 | .02 | .14 | -.36 | .19 | 4.2 |
| Blocks 2 and 3 | .001 | .005 | .22 | .15 | .56 | .03 | 4.0 |

---

[26] The cumulative spikiness was calculated for the mapped physiological output (MPO) measure only, since a similar analysis of the muscle activity measure yielded no informative or interpretable results.

| | | | | | | | |
|---|---|---|---|---|---|---|---|
| Blocks 1 and 3 | .25 | .02 | .001 | .02 | -.56 | .42 | .41 |
| **Condition 3** | | | | | | | |
| Blocks 1 and 2 | .06 | .001 | .004 | -.02 | .47 | .50 | -.79 |
| Blocks 2 and 3 | .42 | .001 | .001 | -.03 | .07 | .78 | -.71 |
| Blocks 1 and 3 | .03 | .001 | .001 | .20 | .09 | .57 | 6.2 |
| **Condition 4** | | | | | | | |
| Blocks 1 and 2 | .44 | .001 | .001 | .35 | .57 | .35 | 12.6 |
| Blocks 2 and 3 | .23 | .34 | .34 | .05 | .62 | .57 | 1.6 |
| Blocks 1 and 3 | .25 | .001 | .001 | .02 | .11 | .13 | .54 |

*5.2.6 Cross-talk between muscles involved in behavior.* To see how the two muscle measurements contribute to the observed behavior across a given set of trials, the mean of the two muscle measurements across an experimental block and an entire experimental condition were compared between the triceps brachii (TB) and flexor carpi radialis (FCR). The results of this analysis are shown in Table 4.

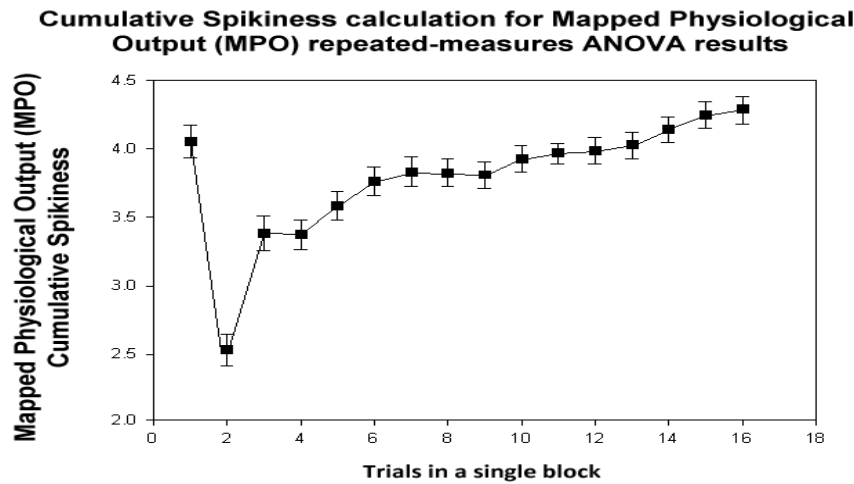

**Figure 8. Significant results for trials, 12x16 analysis of MPO cumulative spikiness.**

*5.2.7 Relationship between muscle measurements and behavioral measurement.* Finally, a correlational analysis was conducted to assess the relationship between the activity of the recorded muscles and the behavioral measurement of physiological output. The results of this analysis are shown in Table 5.

**Table 4. Correlation coefficients between muscle measurements (TB = triceps brachii, FCR = flexor carpii radialis). All correlation coefficients > .40 and < -.40 highlighted in red.**

| | TB vs. FCR Condition 1 | TB vs. FCR Condition 2 | TB vs. FCR Condition 3 | TB vs. FCR Condition 4 |
|---|---|---|---|---|
| **Learning block 1** | -.375 | .009 | -.390 | .409 |
| **Learning block 2** | -.074 | .278 | .129 | .551 |
| **Learning block 3** | -.018 | .226 | -.037 | .323 |
| **block total** | -.064 | -.153 | -.215 | .399 |

**Table 5. Correlation coefficients between mapped physiological output (MPO) and muscle measurements (TB = triceps brachii, FCR = flexor carpi radialis).**

|  | Condition 1 | | Condition 2 | | Condition 3 | | Condition 4 | |
|---|---|---|---|---|---|---|---|---|
|  | TB vs. MPO | FCR vs. MPO | TB vs. MPO | FCR vs. MPO | TB vs. MPO | FCR vs. MPO | TB vs. MPO | FCR vs. MPO |
| **Learning block 1** | -.061 | .011 | .134 | -.004 | .028 | .089 | .033 | .041 |
| **Learning block 2** | -.040 | -.106 | -.138 | -.07 | -.031 | -.029 | -.016 | -.003 |
| **Learning block 3** | .031 | -.169 | .149 | .06 | .034 | -.052 | -.008 | .045 |
| **block total** | -.025 | -.163 | .024 | -.047 | .030 | -.032 | -.003 | .022 |

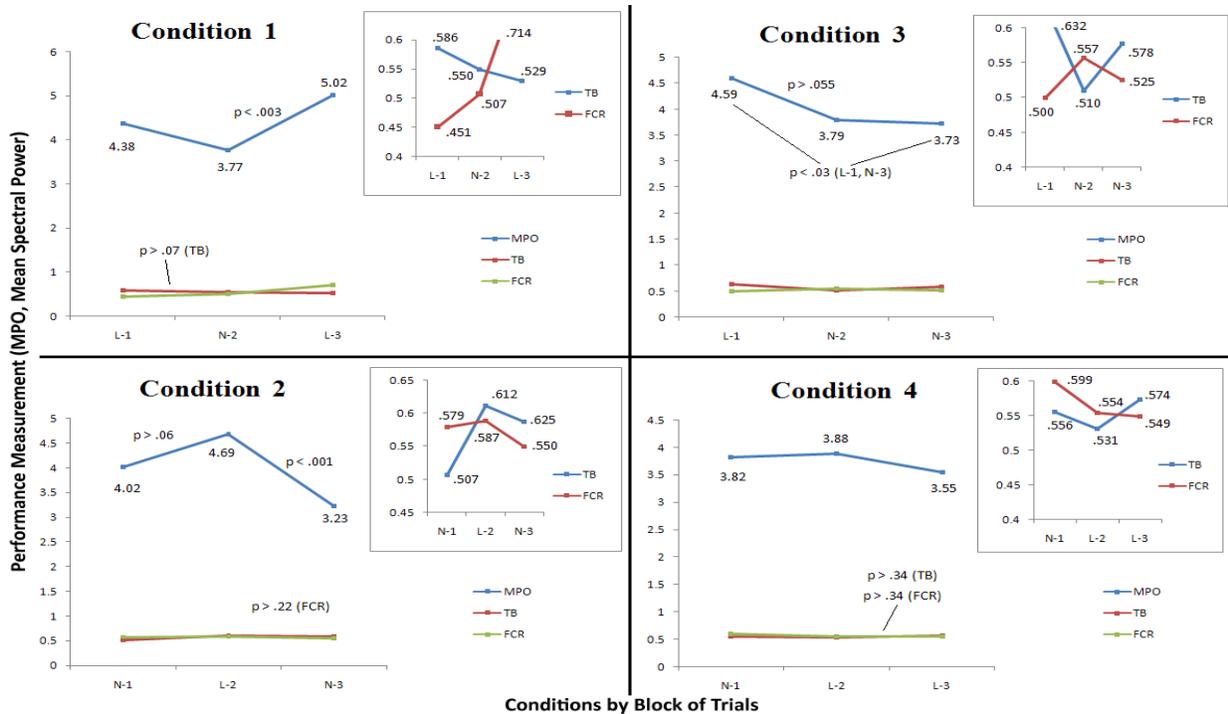

**Figure 9. Results of paired t-tests and means for each measure by condition and block. Significances, near-significances, and means for each condition/block combination are shown.**

*5.2.8   Specific example of the sensitivity analysis.* One lesson learned from the sensitivity analysis is that removing specific datapoints can change the correlation coefficient much more drastically than removing a relatively large number. This may suggest that performance scaling is "critically organized", with certain individuals not conforming to the same behavioral and physiological parameters characteristic of most other members of a population. To demonstrate this graphically, a sample analysis was conducted and shown in Figure 10 (as part of the robustness analysis).

**5.3   Robustness Analysis.** One theoretical claim made here is that robustness of physiological state plays a role in the adaptive response to a perturbatory stimulus. To better understand the role of robustness in this system, I used the fitness calculation for the MPO measurement to determine what the effects of perturbation are on performance before, after, and well after the perturbation is delivered. The results of this analysis are shown in Figure 11.

*5.3.1 Details of robustness analysis.* The first step was to compare trials during which reintroduction of the non-perturbatory stimulus took place with trials during which learning block 3 of the non-perturbatory stimulus took place (see Figure 11). This is slightly different for conditions 1 and 2 than it is for conditions 3 and 4. For conditions 1 and 2, this involves comparing the learning blocks 1 and 3. For conditions 3 and 4, this involves a comparison of learning blocks 2 and 3. This also means that conditions 3 and 4 serve as a control; in these conditions, the perturbation is delivered before the blocks in question.

For each condition, a correlation coefficient and regression coefficient were calculated. These statistics can be contrasted with a visual inspection of the two functions in Figure 11. By combining these two means of analysis, the effects of delivering a perturbation during the learning block 1 becomes evident. For example, there is a systematic fitness dropoff in learning block 3 for condition 4 that is absent in the other three conditions. In addition, condition 3 yields very low correlation and regression coefficients. This also suggests that there is an effect on robustness well after the initial perturbation is delivered; whether or not this is always deleterious is not clear from these data.

*5.3.2   Interpreting results of the robustness analysis.* It is also apparent that delivering a perturbation in the middle of a three phase experimental session may not have much initial effect on fitness. In Figure 10, there is relatively little difference between the two functions, and the correlation and regression coefficients are relatively high. By combining these findings with the paired t-test results, it could be argued that delivering a perturbation after an initial learning block serves as a mechanism for maintaining homeostasis, while delivering a perturbation during learning block 2 may serve as a mechanism for initiating allostatic drive.

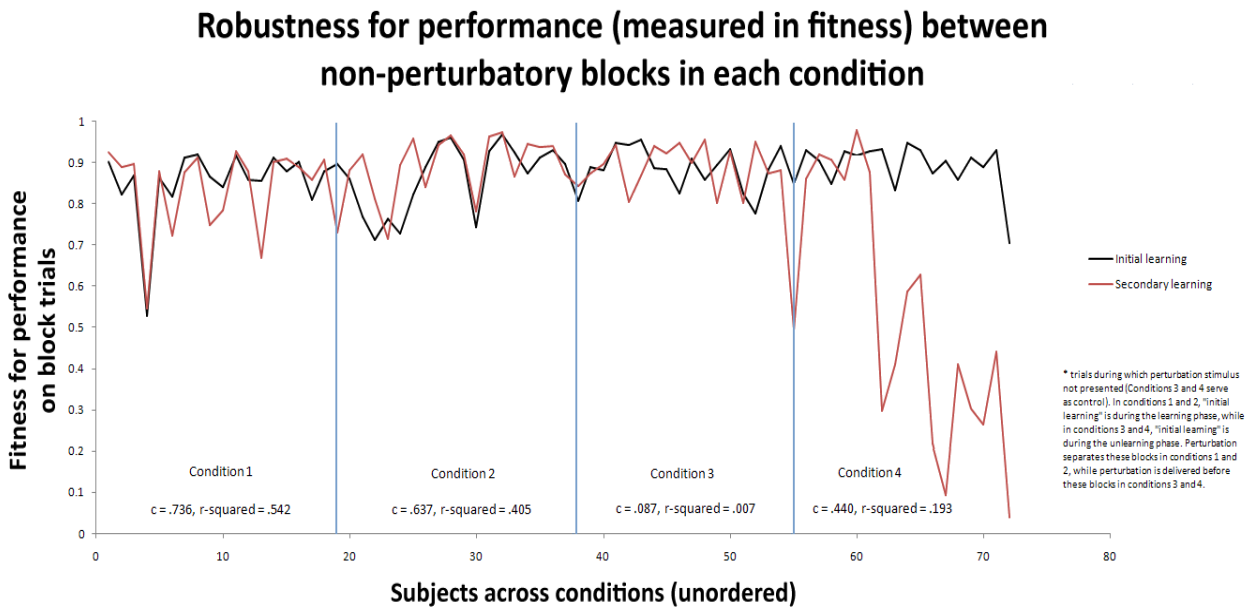

**Figure 10. Robustness analysis for all non-perturbation trials in each of the four experimental conditions.**

**5.4   Data Interpretation: Two Instances of Adaptability of Internal State.** The difference between homeostatic and allostatic regulation involves the degree of robustness exhibited by a physiological system in response to a perturbation. Two examples from the experimental data will be demonstrated here: the trend of performance level across blocks of trials in a single experimental condition, and performance scaling for mapped physiological output and performance scaling (see Figure 11).

For the blue functions in frames A and B of Figure 11, the stability of a single performance level value before, after, or well after perturbation is representative of homeostasis and a robustness to perturbation. Specifically, a similar level of performance should be observed for all non-perturbation blocks in a certain condition. This should hold true whether the perturbation is delivered during either the learning block 2 (block 2 in frame A) or learning block 1 (block 1 in frame B). These results suggest that internal mechanisms such as neural or muscular plasticity act to recover performance on the original force. By contrast, allostatic drive is demonstrated by the red functions in frames A and B. In these cases, performance level is affected by the perturbation either directly or indirectly. Therefore, it can be said that the internal mechanism moves to a new state for that specific set of force stimuli (i.e. naked or loaded prosthetic).

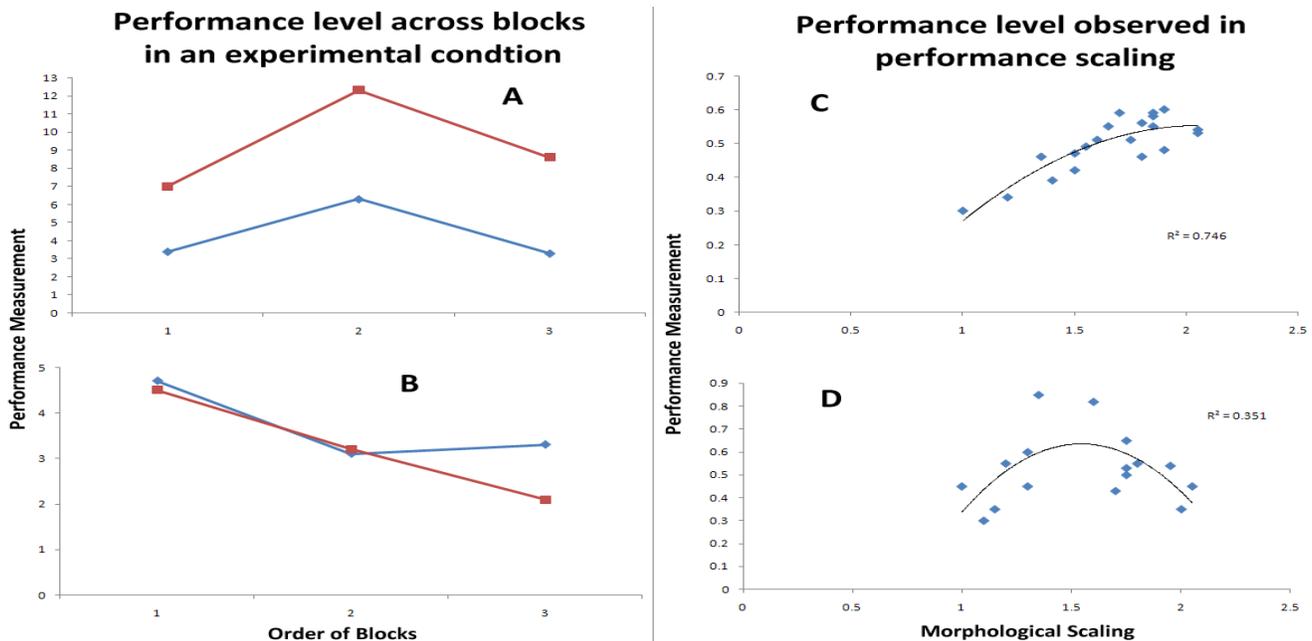

**Figure 11. Two examples of the difference between homeostasis and allostasis. Frame A) homeostasis (blue function) and allostasis (red function) observed in performance level across blocks in an experimental condition when perturbation delivered during block 2, B) homeostasis (blue function) and allostasis (red function) observed in performance level across blocks in an experimental condition when perturbation delivered during block 1, C) homeostasis observed in performance scaling when perturbation delivered during block 1, D) allostasis observed in performance scaling when perturbation delivered during block 2.**

In the case of frames C and D, performance scalings with a higher initial regression coefficient and require the removal of fewer datapoints in the sensitivity analysis (shown in frame C) are likewise representative of homeostasis and provide an additional signature of robustness. Frame D is thus representative of allostatic drive, and is produced by the same mechanisms as those that produce the red functions in frames A and B.

**5.5 Data Interpretation: Range Reduction method for hypo- and hyper-allometry.** To better understand the magnitude of effects between arm morphology, physiology, and behavior, I conducted a set of computational experiments on the data. These experiments were motivated by an observation that the range of each variable contributed to the shape of the function. In addition, there were concerns regarding whether specific blocks produced hyper- or hypo-allometric effects. The size and shape of the humerus and forearm were computed using several measurements[27]. Each approximation of volume was then scaled against each other. The regression equation and scaling factors (i.e. $b_1$ and $b_2$) were calculated for these relationships, in addition to a range-based normalization of the function shape (see footnote 39). Results of this analysis are shown in Table 6.

*5.5.1 Range-reduction calculation of hyper- and hypo-allometry.* To normalize the effects of reducing the range of one or both axes, a measurement was used to capture the shape of the nonlinear regression function[28]. When the value of this measurement is 0, it is roughly equivalent to a monotonic linear function. Positive values indicate an increase in the range of the y-axis relative to the x-axis (hyper-allometry), while negative values indicate an increase in the range of the x-axis relative to the y-axis (hypo-allometry). To achieve this, three parameters are defined: $\alpha$, $\beta$, and $\Omega$. Alpha is the range of data on the x-axis (which is the maximum value subtracted by the minimum value), beta is the range of data on the y-axis, and omega is the reduction of the range of data on either or both axes by some factor. The schematic in Figure 13 describes these relationships in graphical form.

**Table 6. Statistics related to performance scaling regression by approximation of arm anatomy, experimental block, and performance measurement.**

| Criterion | Measure | Block | Regression Coefficient | $b_1$ | $b_2$ | Hypo/Hyper-allometry (range-based method) * |
|-----------|---------|-------|------------------------|-------|-------|----------------------------------------------|
| Size | MPO | 1 | 0.165 | 4.583 | -8.243 | 1.65 |
| Size | MPO | 2 | 0.374 | 2.179 | -5.204 | 1.55 |
| Size | MPO | 3 | 0.197 | 4.424 | 5.127 | 1.73 |
| Shape | MPO | 1 | 0.041 | 0.975 | -1.65 | -0.18 |
| Shape | MPO | 2 | 0.1 | 0.209 | -0.617 | -0.22 |
| Shape | MPO | 3 | 0.083 | 0.875 | -1.612 | -0.16 |
| Size | TB | 1 | 0.102 | -14.05 | 6.147 | 2.87 |
| Size | TB | 2 | 0.099 | 4.487 | -2.95 | 1.66 |
| Size | TB | 3 | 0.045 | 6.097 | -3.175 | 2.69 |
| Shape | TB | 1 | 0.054 | -3.152 | 1.355 | 0.19 |
| Shape | TB | 2 | 0.037 | 0.863 | -0.563 | -0.18 |
| Shape | TB | 3 | 0.167 | 4.222 | -2.069 | 0.14 |
| Size | FCR | 1 | 0.096 | 12.12 | -5.537 | 2.91 |
| Size | FCR | 2 | 0.064 | 2.177 | -1.764 | 3.07 |
| Size | FCR | 3 | 0.124 | 5.187 | 3.589 | 2.13 |
| Shape | FCR | 1 | 0.097 | 3.772 | -1.737 | 0.20 |
| Shape | FCR | 2 | 0.135 | 0.975 | -0.791 | 0.25 |
| Shape | FCR | 3 | 0.145 | 1.905 | -1.276 | -0.03 |

**\* value for $\Omega$ in hypo/hyper-allometry calculation held constant at 1 across all comparisons.**

*5.5.2 Specifics of range reduction method.* To test the robustness of these specific performance-morphology relationships, a range of potential performance and anatomical resamplings were conducted[29]. For the mapped physiological output measurement, the following question was asked: if the range of phenotypes or performance responses to perturbations of greater magnitudes are expanded or reduced, what would the effect be on the regression equation? In simpler terms, changes in magnitude for either of these factors were expected to enhance the degree of hyper- or hypo-allometry in this scaling relationship.

---

[27] The size scaling was calculated by comparing the volume of two cylinders: one representing the forearm, and the other representing the humerus. Volume was the product of the circumference and length of each segment. The shape scaling was calculated by dividing the height (in this case the diameter of the segment) by the width for both segments. For symmetrical shapes such as a square, this results in a value of 1.

[28] The range-based calculation is RBC = [ $(\alpha)$ * $(\Omega_x)$ ] / $(\beta)$ * $(\Omega_y)$ ] − 1. When the range of both the x- and y-axes are equal, is results in a value of 1. Subtracting 1 from this value results in 0, which represents a "neutral" nonlinear function.

[29] The purpose of the range-based method is to assess hypothetical growth and form scenarios in development and evolution. While done in an abstract manner, the range-based method identifies potential constraints that could affect

Figures 13 and 14 demonstrate the range-based method measurement value for a range of scaling and performance measurement reductions. In Figure 12, mapped physiological output and the size (blue) and shape (red) scalings were reduced by 25 (.25), 50 (.5), and 75 (.75) percent. The values on both axes were also held constant (at a value of 1.0).

The manipulations demonstrate how changes in each variable may affect interactivity in a number of potential settings. For example, reducing the performance measurement while keeping the morphological scaling constant results in a large increase in the range-based measurement, and indicates significant hyper-allometry. When the opposite manipulation is performed (reducing the morphological scaling while keeping the performance measurement constant), the range-based measurement tends towards hypo-allometry.

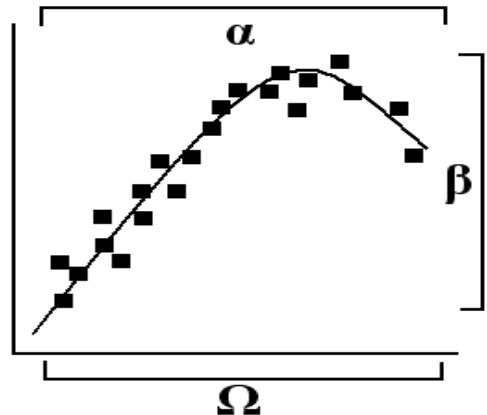

**Figure 12. Schematic of range-based calculation for hypo- and hyper-allometry. Three parameters for derivation of calculation: α is the range along the x-axis (scaling), β is the range along the y-axis (performance measurement), and Ω is the artificial scaling of the data by some factor. NOTE: the parameter Ω can refer to the x-axis, the y-axis, or both. Shown here is the parameter Ω with a value of 1.0 for the x-axis.**

Figure 14 shows the same set of experiments conducted on the muscle measurements. Triceps brachii (TB) activity, flexor carpi radialis (FCR) activity, and the size (blue) and shape (red) scalings were reduced by 50 (.5) percent. As in the case of mapped physiological output, the values on both axes were also held constant (1).

As in the case of mapped physiological output, reducing the performance measurement by 50 percent increases the range-based measurement by a similar magnitude for both muscles. Also similar to what is observed with the mapped physiological output performance measurement, the size scaling shows a larger effect than the shape scaling.

**5.6 Hypothesis-based interpretation of data.** There are numerous lessons to be learned from the patterns revealed herein. These are both related to explicitly stated hypotheses and more general regulatory principles. The lessons directly related to the hypotheses can be summarized as follows:

$H_A$: Augmentation by Subsequent Effects. Switching induced by alternating technological environments has an effect on performance. This effect unfolds across multiple blocks of experimental trials.

$H_B$: Environmental Information Hypothesis. Environmental perturbation with the prosthetic device results in increased variability of performance level measurements such as muscle activity and mapped physiological output, and contributes to allostatic drive.

$H_C$: Functional Static Allometry Hypothesis. The bounds of adaptability, which is represented by the size and shape of the adult phenotype, has an implicit effect upon performance level expressed during and after augmentation.

$H_D$: Allostatic Drive Hypothesis. Switching between technological environments has an effect on the baseline as measured in between blocks of trials.

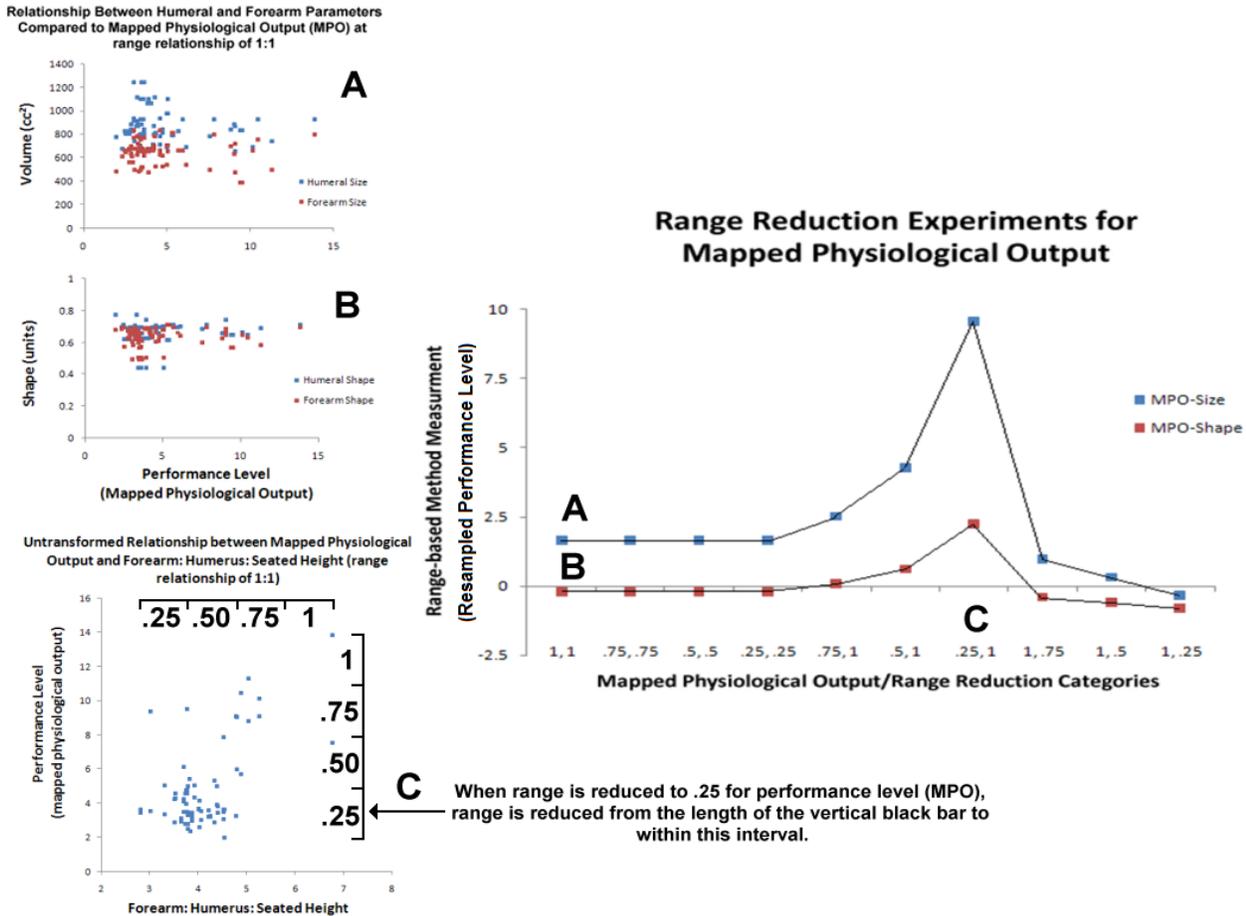

**Figure 13. Results for range reduction experiments conducted on mapped physiological output data (range-based measurement averaged across experimental blocks). Main graph is displayed on right-hand side. Reduction categories are listed as *m,n*, where *m* is reduction/enlargement of performance measure, and n is reduction/enlargement of the allometric scaling. A and B represent the normal range of size and shape vs. performance level before range reduction. C is the relationship between performance scaling and performance level before range reduction.**

*5.6.1 Not all forms of switching are equal.* The first of these lessons, which arises from tests of Hypotheses A and B, is that not all forms and types of switching are equal. This can be seen in the paired t-test and ANOVA results. One example is that there is a reasonably strong effect of switching measured using mapped physiological output when a perturbation is introduced during learning block 2 (Conditions 1 and 2). That this is true for a differential measure of force production suggests that introducing a force perturbation sets up changes in performance level later on in the session. The key seems to be the contrasting nature of the two treatments: a wider range of previous experience can push the entire system in a given direction.

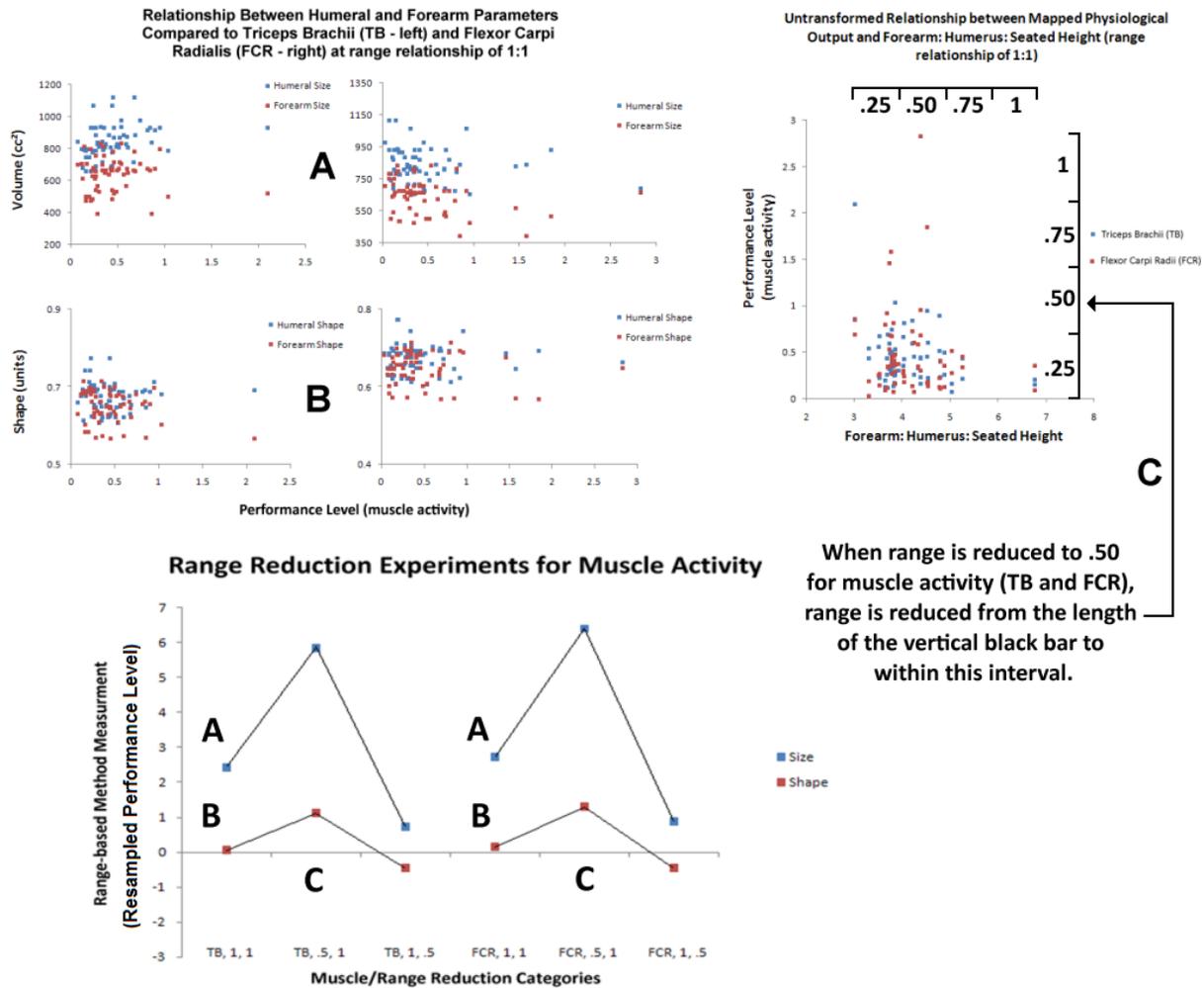

**Figure 14. Results for range reduction experiments conducted on muscle activity data. TB = triceps brachii, FCR = flexor carpi radialis (range-based measurement averaged across experimental blocks). Main graph is displayed at bottom. Reduction categories are listed as *m,n*, where *m* is reduction/enlargement of performance measure, and n is reduction/enlargement of the allometric scaling. A and B represent the normal range of size and shape vs. performance level before range reduction. C is the relationship between performance scaling and performance level before range reduction.**

Secondly, switching from a loaded to a naked controller when the perturbation is introduced at learning (Condition 3) is nearly statistically significant. This, combined with the further effect of maintaining the naked controller in learning block 3 (Condition 3), again suggests that a force perturbation sets up changes in performance level later on in the condition. In this case, introducing a large-scale force distortion (which the loaded controller provides) as the initial condition widens the range of experience more than a smaller-scale distortion.

*5.6.2 Data trends reveal multiple regulatory mechanisms at work.* The trends seen in muscle activity, which is a test of Hypothesis B, are hard to get at using statistical significances. Neither the ANOVA nor the t-test results were particularly revealing. Correlational analyses conducted between muscles within single blocks of trials (Tables 8 and 9) revealed that muscles in the forearm and humerus do not work together most of the time during the experimental activity or during perturbation. The only exception might be during Condition 4, where a positive correlation above .40 exists immediately during and after perturbation for the naked controller during the learning block of trials and the loaded controller during the learning block 2, respectively. Interestingly, the mapped physiological output for both of these blocks of trials is constant, with a decrease coming only in the learning block 3 of Condition 4. This suggests that when muscles work together by contracting at similar frequencies, force output by the muscles and error detection by the brain's internal model is stabilized. In addition, the cumulative spikiness transformation of the mapped physiological output reveals that greater variability is exhibited in controlling force output. As shown in Figure 7, this measure decreases sharply during the second trial, and rises sharply thereafter. This could be due to a common immediate response to conditions of the technological environmental, while over time there is a divergent response to the specific treatment across individuals.

While differences exist across individuals for variability across trials within a block, common patterns of performance across individuals occur between different blocks of trials. The covariance of mapped physiological output between blocks of trials demonstrates that a strong positive linear dependence exists between during and well-after perturbation (learning blocks 1 and 3) in Condition 3 and during and immediately after perturbation (learning blocks 1 and 2) in Condition 4. This suggests that among all individuals, force output control is relatively unaffected after a loaded controller perturbation. The same does not appear to be true after a naked controller perturbation, especially after considering the results for Condition 1. However, given enough time, force output control returns to what it was during the perturbation. This may be not only be related to the three stages of memory formation discussed earlier, but also to the three stages of long-term changes at the synaptic level.

Visual inspections of the data reveal that the muscle activity value trends in different directions for each muscle and experimental condition. The muscle representing the forearm (flexor carpi radialis) exhibited elevated muscle activity during perturbation when switching occurs during learning block 2 (Conditions 1 and 2). This elevation of activity was even more pronounced when the loaded controller was presented during the learning block 3 (Condition 1). By contrast, muscle activity decreases when a loaded controller is introduced during learning block 3. This results in an overall decrease in muscle activity across Condition 2. When switching occurs during the learning block (Conditions 3 and 4), two distinct patterns of activity are observed. Activity in Condition 3 shows a pattern similar to that of Condition 2, except there is an overall increase in activity across the condition. In Condition 4, muscle activity decreases after perturbation, but is maintained as the technological environment is maintained. For the muscle representing the humerus (triceps brachii), perturbation produced a suppressive effect across Conditions 1 and 3 and an enhanced effect across Conditions 2 and 4. In Conditions 2 and 3, the perturbation itself has an enhanced effect, whereas removing the perturbation has a suppressive effect in Conditions 1 and 4.

*5.6.3   Two general mechanisms for adaptability and regulation.* There appear to be two mechanisms at work across each condition. One is the adaptive ratchet mechanism, which explains a V-like pattern in performance level values between learning blocks 1 and 3. This addresses the predictions made by Hypothesis D, and explains larger-scale changes during or immediately after the perturbation which drive smaller-scale changes over the duration of the block. Challenging the system with a perturbation forces the performance level to increase or decrease sharply either when the perturbation is introduced or right after it is removed. This increase or decrease is countered during learning block 3, but results in an overall increase or decrease in performance level for that condition.

As revealed in the last section, the data show two general patterns: a V-shaped pattern of change in muscle activity and mapped physiological output, and a recovery of the muscle activity and mapped physiological output attained either before or when the perturbation was introduced. These patterns can be described in terms of mechanisms: the former is an example of allostatic drive initiated by an adaptive ratchet mechanism, while the latter is an example of a robustness mechanism which maintains homeostasis. Robustness is a mechanism that allows the human system to absorb the introduction of a perturbation. Perhaps this can best be understood in terms of what the muscle activity measurements mean. Changes in the muscle activity power spectrum are referred to as spectral shifts (Duchene and Goubel, 1990). When induced using a series of isometric contractions and a constant torque stimulus, shifts to lower frequencies often indicate muscle fatigue. In the case of a dynamic stimulus, shifts to lower muscle activity values may indicate a reliance on morphological rather than neural control. In the case of allostatic drive, an inverted V-shaped pattern indicates morphological compensation (e.g. contributions of muscle and limb geometry) is responsible for the changes in performance, while the standard V-shaped pattern indicates neural compensation predominates. The robustness mechanism may become active when morphology and the neural mechanisms contribute equally to the perturbation response.

At the anatomical and molecular scale, ultrastructural changes (see Palacios-Pru and Colasante, 2004) at the neuromuscular junction and perhaps even synapses in selected brain regions may be the culprit. Palacios-Pru and Colasante (2004) experimented with a guppy (*Lebistes reticulates*) model where variations in water current were used as a perturbatory force stimulus. Changes in the amount of neurotransmitter expressed and vesicular density were transient when the perturbation was introduced once and then removed. These same changes were more permanent when the interval between training session or perturbations were reduced. In short, repeated and interspersed presentations of a significant force stimulus can induce adaptations that can have a positive effect on performance.

The second mechanism appears to be a form of robustness which acts to maintain the performance level regardless of perturbation. Examples of which are the triceps brachii and flexor carpi radialis muscle activity trends across Condition 1, and the mapped physiological output trends across Conditions 3 and 4. In all three cases, neither introducing nor removing the perturbation results in an immediate effect. In the case of mapped physiological output in Condition 3, removing the perturbation has a slight effect, while maintaining the same technological environment affects the performance level in the same way albeit in a less pronounced fashion.

At the anatomical and molecular scale, the changes responsible for setting up this robustness mechanism may be the same changes associated with environmental enrichment. In a rat model, it has been shown by Chang and Greenough (1984) that increasing the amount of sensory stimulation in an environmental setting is associated with the upregulation of mRNA (transcriptional) production and growth factors that have a positive effect on future performance.

*5.6.4 Range-reduction and scaling sensitivity.* One test of Hypothesis C involves the scaling sensitivity analysis and hypo- and hyper-allometry range-reduction exercises that were conducted in sections 5.5.1 ad 5.5.2. One insight this exercise provides are the relative contributions of morphological control, neural control, and muscular control to adaptive responses. For example, differential effects found for various scaling in the sensitivity analysis suggest a non-statistical interaction between neural and morphological control. In cases where only a few individuals needed to be removed from the dataset to yield a reasonable regression coefficient, it could be that morphological control for that scaling is very strong. In other words, certain features of morphology conform to the demands of performance across a range of conditions. In cases where many individuals needed to be removed, support for improving performance across a range of conditions has been shifted by evolution to allelic differences between subpopulations. In these cases, the individuals thrown out of the analysis may constitute carriers of genotype A, while individuals that remain may constitute carriers of genotype B.

In addition to understanding the contribution of each type of control, different regulatory mechanisms can be associated with the various types of control. For example, neural control is driven by an internal model which produces a ratchet-like mechanism for adaptation. In this case, a perturbation provides the means to explore a range of responses which loosely correspond to the physiological range of the individual. After this initial perturbation, the system finally settles into a stable state which will differ from the initial condition, but not as much as the response during the perturbation. Changes driven by muscular control may involve a robustness mechanism which provides a mechanism for adaptation that is semi-independent of the perturbation. The robustness analysis and muscle activity baseline ANOVA analysis are tests of Hypothesis B and D, and reveal that this mechanism affects both fitnesses exhibited across the population and baseline muscle activity. Muscle fiber recruitment and protein expression in the cross bridges of the muscle fibers during contraction may allow for this robustness mechanism to compensate for morphological control and override neural control under certain conditions (Rome et al, 1999).

The far right column in Table 15 provides a range of values that determine how much morphological proportions and performance measures vary when the transformations are not applied. The greatest degree of hyper-allometry is exhibited when the muscle activity of triceps brachii and flexor carpi radialis is scaled against the humerus-to-forearm size measurement. In Figures 15 and 16, a factor between .25 and 1 were used for both variables. For mapped physiological output, the greatest effect of hyper-allometry is when performance is reduced by a factor of .25 and the allometric scaling is not reduced. For both muscles, the greatest effect is when the performance measure is reduced by a factor of .50 and the allometic scaling is not reduced. This true for both size and shape measurements. All this is an indirect test of Hypotheses C and D, and speaks to how adaptable the physiological mechanisms are under the experimental conditions, especially when the size and shape of the phenotype are held constant.

*5.6.5 Shifts in MPO and muscle activity.* There are two interpretations for shifts in MPO and muscle activity during the course of the experiment. Recall that there are two general effects: movement to a new physiological state, and recovery of the original physiological indicators before perturbation. Furthermore, there are two possible secondary effects: one when the perturbation has no significant effect on performance, and the other when a significant difference occurs between two non-adjacent blocks (such as before and after a perturbation or during and well-after a perturbation).

All of these results can be explained in terms of the mapping between the virtual environment and physiological system driving action within. For example, when there is recovery of the original level of performance in the MPO measurement, the mapped physiological output retains its mapping to both the virtual activity and the controller device. When the MPO value is low or decreases, the combination of environmental conditions is more isomorphic to the current state of physiological regulation. When the value for MPO is high or increases, the inverse is true.

A similar range of scenarios is true for the muscle activity measurement. Over sustained periods of activity, muscle naturally fatigues (Duchene and Goubel, 1990). During dynamic stimulation, the mechanical state of the muscle also changes. When the muscle activity measurement undergoes spectral shifts, it represents differences in the discharge rate of muscle fibers (Barry and Enoka, 2007). This is a compensatory mechanism in muscle that is a signature of metabolic and mechanical changes that occur while the physiological system attempts to match immediate environmental demands (Barry and Enoka, 2007).

*5.6.6 Explicit contributions of homeostasis and allostatic drive to regulation*. Finally, these regulatory mechanisms can be discussed more explicitly in terms of homeostasis and allostatic drive. In Figure 12, homeostasis can be defined by a performance level that remains stable during the time period (or block of trials) when a perturbation is not present. When the perturbation is introduced during the learning block, performance level decreases after the perturbation and then remains level during learning block 3. When the perturbation is introduced during learning block 2, performance level takes on a V shape across the condition but does not result in a net increase. In the same figure, allostatic drive can be defined as either a net increase or decrease across the condition. This can either take the form of a continuously increasing or decreasing function, or a change produced by the adaptive ratchet mechanism.

Homeostasis and allostatic drive can also be described in terms of performance allometry. In the case of homeostasis, the polynomial function exhibits a cane-shaped trend. In the case of allostatic drive, the polynomial function exhibits more of a U-shaped trend. The reason for this difference may be due to an increase in variability of performance at any particular scaling due to the adaptive ratchet mechanism. By contrast, homeostasis results in a performance level measurement that is fixed relative to a given scaling, and more accurately reflects the contributions of morphological control. This seems to result in an optimal scaling/performance level equilibrium point that is closer to the maximum, whereas this optimal point moves towards the center of the range in cases of allostatic drive.

While the position of this optimum point along the range of the function needs to be investigated further, it may be related to a static form of hyper- and hypo-allometry driven by changes in the performance level measure. According to the range-based simulation experiment in 5.6.1, hyper- and hypo-allometry were induced in the nonlinear function by transforming the range of performance and morphological measures according to a pre-defined factor. This tells us how the optimal point changes as the scaling of the humerus-to-forearm size and shape and performance level as measured by mapped physiological output and muscle activity shrink or grow larger, essentially extrapolating the nonlinear function derived from real data to a larger range of potential values.

*5.6.7 Determining the roles of homostatic regulation and allostatic drive*. One interesting result of these experiments was that a perturbation, whether it was of a different type or at a different position in the condition, has a different effect on physiological output and muscle activity. This can be explained theoretically by further considering the mechanism of internal regulation and how they relate to the capacity for adaptation. Table 7 demonstrates how homeostatic mechanisms differ from allostatic mechanisms.

Sterling and Eyer (1988) have proposed a difference between homeostasis and allostasis. According to their view, homeostatic control involves a single feedback loop and relatively simple regulatory mechanism, while an allostatic system involves multiple regulatory mechanisms. These often interacting mechanisms respond to a broader range of stimuli. This can be understood by comparing how much fitness is retained for the mapped physiological output measurement for blocks that immediately followed a perturbation and those that are well-after a perturbation (all in comparison to an initial learning block, as was done in the robustness analysis). Blocks in which a loaded controller is used which are also well-after a naked controller perturbation retain 30.6% of fitness across subjects. By contrast, blocks in which a naked controller is used which are also after a loaded controller perturbation retain 103.7% of fitness across subjects. Blocks which involved a loaded controller after a naked controller perturbation and blocks which involved a naked controller well-after a loaded controller perturbation retained 97% and 95.9% of fitness, respectively. In the case where 30.6% of fitness was retained, it could be that multiple and even indirect regulatory mechanisms are at play. By contrast, the case where 103.7% of fitness was maintained might involve a strongly robust mechanism with a relatively simple regulatory structure.

**Table 7. Major differences between Homeostasis and Allostatic Drive**

|  | Homeostasis | Allostatic Drive |
|---|---|---|
| Systemic response | Robustness | Brittleness |
| Potential for compensation, regulatory response | One compensatory mechanism (responds to challenge of perturbation – learning-related adaptation in nervous system) | Many compensatory mechanism (may force peripheral tissues to adapt, Initial response to perturbation unstable) |
| Effects of standing variation | Allelic factors, gene expression pathways allow for compensatory mechanism, limits further adaptability | Allelic factors, gene expression pathways may exist for multiple stable states, no immediate limit on further adaptability. |
| Extent of learning-related changes (physiological) | Extensive, long-term changes in synaptic efficacy (more global consolidation) | Limited changes in synaptic efficacy (lack of global consolidation) |
| Cause of systemic response (perturbatory stimulus) | Device with familiar shape, operating kinematics | Device with novel shape, requires novel operating kinematics |
| Performance scaling | "Core" variation exhibited | "Extreme" variation exhibited |
| Adaptive potential of system | Overall adaptability low, adaptability of specific compensatory mechanism high | Overall adaptability high, adaptability of specific compensatory mechanism high |

In addition, this result shows that there may be two separate regulatory mechanisms: one governing the retention of temporal information about the perturbation, and another that compares afferent force information with force information obtained during the perturbation. In the case of homeostatic responses, this information is most likely used to produce a neuromuscular response within the bounds of current physiological capacity. In the case of conditions that initiate allostatic drive, this information triggers molecular pathways in cell populations that induce adaptive tissue responses.

## 6.0   Conceptual Outcomes

Overall, perturbation of the environment using a computational system should induce changes in behavior and physiology. These changes are related to learning and memory consolidation, and result in a set of activity-dependent adaptations which optimize performance. These adaptations might conspire with physiological changes associated with previous learning and morphological features to either further augment performance of constrain performance gains. One conceptual outcome from this experimental paradigm involves treating changes imposed by performance augmentation probes as phasic phenomena. Further interpretation suggests that behavioral and physiological indicators exhibit homeostasis before and after training, and that this relative stability represents stable states that can be achieved though environmental perturbation. Figure 16 shows a graphical representation of this scenario.

The research questions posed earlier provide a means to examine the relationship between homeostasis and adaptation in a systematic manner. Specifically, homeostasis and adaptation can be conceptualized as a bivariate relationship for a level of performance time-series. In Figure 16, homeostasis as a function of adaptation can be manipulated in two ways. The blue function demonstrates how the values for x and y change during perturbation, allostatic drift, and augmentation. One is to induce change along the x-axis, which translates into magnitudinal perturbation. The other is to induce change along the y-axis, which translates into durational perturbation. This induction occurs as a by-product of learning block 2. The magnitude and duration of learning block 2 response is related to the amount of allostatic drive observed in learning block 3. A third manipulation involves the duration during which the original level of homeostasis was maintained. In terms of both memory traces and muscle activity, the longer a system remains in a particular state, the harder it is to change with learning block 2 or a similar form of environmental perturbation.

In addition, some perturbations are more effective than others. In Figure 15, two cases (red and blue function) are demonstrated. While reaching behavior portrayed by both functions are both subject to perturbation and start off at a similar stable state, only one function demonstrates conditions that serve as a trigger for allostatic drive. As a result, while the blue function clearly shows evidence of augmentation, while the red function does not.

It is also of note that in contexts outside of this experiment, physiological changes based on prior learning of similar activities can affect how the memory trace is consolidated and the physiological baseline is affected. Ultimately, this affects how much the physiological range is modified during augmentation. If the overlap in physiological changes required or transfer between similar activities is significant, the degree of allostatic drive is limited given a specific amount of perturbation. In the conceptual result, this will result in a smaller $\Delta y$ given a larger $\Delta x$. Likewise, if there is little contribution from previously learned activities, a greater change in the level of performance would occur given a relatively smaller amount of perturbation over time.

Finally, there is a longer-term aspect to the relationship between allostatic drive and how the environment is perturbed over time. Recall that homeostasis is considered to be a stable phase that is modified by either repeated or sustained environmental perturbations. Over time, these perturbations may induce plasticity mechanisms which act to produce a permanent adaptation that causes the homeostatic phase to drift to a new dynamic range. This is hard to uncover given current experimental limitations. However, future investigations of modeling the adaptability and evolvability of sensorimotor behavior in humans and operation of the internal model may provide answers.

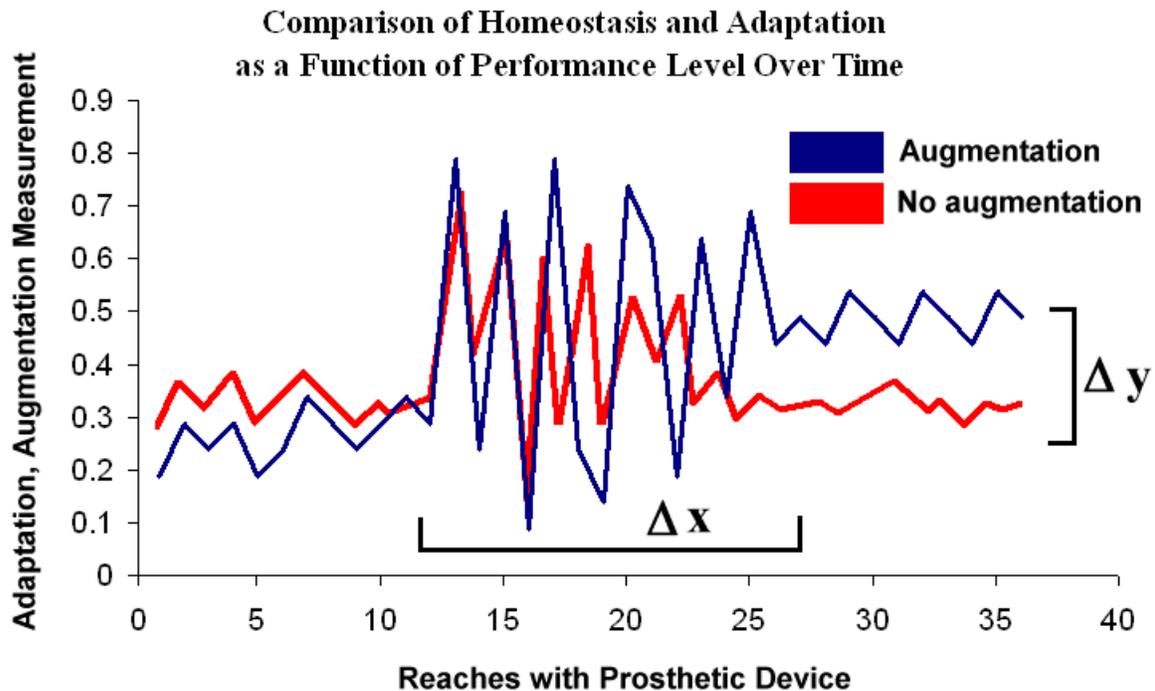

**Figure 15. Graphical representation of how environmental perturbation may drive adaptation represented by changes in the system parameters that accompany allostatic drive.**

### 7.0    Contributions.

Several interesting points can be raised from this work that may be useful for the theoretical underpinnings of performance augmentation. One of them involves the nature of internal states responsible for these behaviors and different ways by which these mechanisms can be modeled. In this paper, a network of brain centers were proposed as being responsible partially responsible for producing an adaptive behavioral output. Future augmentation strategies and ways to determine how much this brain network adapts to environmental perturbations may include targeted repetitive transcranial magnetic stimulation (rTMS) and the administration of pharmacological agents (Boniface and Ziemann, 2003 – see Box 6 for additional detail).

<div style="border:1px solid black">

**Box 6. Definition of molecular bases of augmentation.**

In the context of behavior, molecular-level changes can play a role in the effects of augmentation, especially when an environmental mutant (introduced in Section 2.2.2) is created. In particular, behavioral conditioning can affect transcriptional pathways in a way that is both predictable and controllable using non-invasive, behavior-based experimental methods. Using a genetically modified mouse model, Athos et.al (2002) have isolated upregulation of the CRE signaling pathway (one important pathway in memory consolidation) in a specific brain region in response to fear conditioning. While molecular pathways respond both directly to environmental stimuli and during the occurrence of memory consolidation in control physiologies, Sweatt (2003, Page 49) that awareness of

</div>

molecular-level correlates of performance may determine how behavioral stimuli should be delivered to an individual. In the case of augmentation, this might be reflected in the timing and duration of perturbatory stimuli. In the context of initiating performance augmentation, these molecular processes might also be disrupted using behavioral-level perturbations or inactivation technologies such as pharmacological agents or rTMS.

In both cases, inhibitory stimuli delivered directly to specific synapses or brain regions are used to modify inhibitory mechanisms such as cerebellar memory consolidation or intracortical inhibition. These inhibitory actions in turn serve to modify the excitability of spinal motorneurons (Quartarone et.al, 2004), which can in turn act to facilitate performance augmentation. Another of these involves the variational underpinnings of these behaviors; genomic and endocrine system variation inherent to the individual can influence these internal processes. Finally, the way in which physiology is regulated during learning acts to produce an augmented response resulting from the degree of coupling between human and machine.

## References:

Aflalo, T.N. and Graziano, M.S.A. (2006). Possible origins of the complex topographic organization of motor cortex:reduction of a multidimensional space onto a two-dimensional array. Journal of Neuroscience, 26, 6288-6297.

Aoki, F. (1990). Activity patterns of the upper arm muscles in relation to direction of rapid wrist movements in man. Experimental Brain Research, 83(3), 679-682.

Athos, J., Impey, S., Pineda,V.P, Chen, X., and Storm, D.R (2002). Hippocampal CRE-mediated gene expression is required for contextual memory formation. Nature Neuroscience, 5, 1119-1120.

Barry, B.K. and Enoka, R.M. (2007). The neurobiology of muscle fatigue: 15 years later. Integrative and Comparative Biology, 47(4), 465-473.

Bejan, A. and Marden, J.H. (2006). Unifying constructal theory for scale effects in running, swimming, and flying. Journal of Experimental Biology, 209, 238-248.

Boniface, S. and Ziemann, U. (2003). Plasticity in the human nervous system: investigations with transcranial magnetic stimulation. Cambridge University, Cambridge, UK.

Bridgeman, B. (2007). Efference copy and its' limitations. Computers in Biology and Medicine. 37(7), 924-929.

Bridger, R.S. (1995). Introduction to Ergonomics. McGraw-Hill, New York.

Brown, I.E., Scott, S.H., and Loeb, G.E. (1995). 'Preflexes' - programmable high-gain, zero-delay intrinsic response of perturbed musculoskeletal systems. Society for Neuroscience Abstracts, 21, 562.9.

Brunelli, D., Farella, E., Rocchi, L., Dozza, M., Chiari, L., Benini, L. (2006). Bio-feedback system for rehabilitation based on a wireless body area network. IEEE Conference on Pervasive Computing and Communications Workshops, 4, 13-17.

Bundle, M.W. Ernst, C.L. Bellizzi, M.J., Wright, S., and Weyand, P.G. (2006). A metabolic basis for impaired muscle force production and neuromuscular compensation during sprint cycling. American Journal of Physiological Regulation and Integrative Comparative Physiology, 291, R1457–R1464.

Chang, F.L. and Greenough, W.T. (1984). Transient and enduring morphological correlates of synaptic activity and efficacy change in the rat hippocampal slice. Brain Research, 309(1), 35-46.

Cheverud, J.M. (1982). Relationships among ontogenetic, static, and evolutionary allometry. American Journal of Physical Anthropology, 59, 139-149.

Chiari, L., Dozza, M., Cappello, A., Horak, F.B., Macellari, V., Giansanti, D. (2005). Audio-biofeedback for balance

improvement: an accelerometry-based system. IEEE Transactions on Biomedical Engineering, 52(12), 2108-2111.

Chin, H.R. and Moldin, S.O. (2001). Methods in Genomic Neuroscience. CRC Press, Boca Raton, FL.

Clark, A. (2003). Natural-born cyborgs: minds, technologies, and the future of human intelligence. Oxford Press, Oxford, UK.

Collins, S., Ruina, A., Tedrake, R., and Wisse, M. (2005). Efficient Bipedal Robots Based on Passive-Dynamic Walkers. Science, 307, 1082-1085.

Conditt, M.A., Gandolfo, F., and Mussa-Ivaldi, F.A. (1997). The motor system does not learn the dynamics of the arm by rote memorization of past experience. Journal of Neurophysiology, 78(1), 554-560.

Delcomyn, F. (1980). Neural basis of rhythmic behavior in animals. Science, 210, 492-498.

DeMarse, T.B. and Dockendorf, K.P. (2005). Adaptive flight control with living neuronal networks on microelectrode arrays. IEEE International Joint Conference on Neural Networks, 3, 1548-1551.

DeMarse, T.B., Wagenaar, D.A., Blau, A.W., and Potter, S.M. (2001). The Neurally Controlled Animat: biological brains acting with simulated bodies. Autonomous Robots, 11, 305–310.

Dietz, V., Gollhofer, A., Kleiber, M., and Trippel, M. (2004). Regulation of bipedal stance: dependency on 'load' receptors. Experimental Brain Research, 89(1), 229-231.

Donchin, O. and Shadmehr, R. (2004) Change in desired trajectory caused by training in a novel motor task. Proceedings of the IEEE EMBS, 26, 4495-4497.

Duchene, J. and Goubel, F. (1990). EMG spectral shift as an indicator of fatiguability in a heterogeneous muscle group. European Journal of Applied Physiology, 61(1-2), 81-87.

Dworkin, B.R. (1993). Learning and Physiological Regulation. University of Chicago Press, Chicago.

Fagg, A.H., Hatsopoulos, N.G., de Lafuente, V., Moxon, K.A., Nemati, S., Rebesco, J.M., Romo, R., Solla, S.A., Reimer, J., Tkach, D., Pohlmeyer, E.A., and Miller, L.E. (2007). Biomimetic Brain Machine Interfaces for the Control of Movement. The Journal of Neuroscience, 27(44), 11842-11846.

Fellous, J.M. and   Linster, C. Computational models of neuromodulation. Neural Computation, 10(4), 771-805 (1998).

Fernandez-Ruiz, J. and Diaz, R. (1999). Prism Adaptation and Aftereffect: specifying the properties of a procedural memory system. Learning and Memory, 6(1), 47-53.

Flanagan, T.C. and Pandit, A. (2003). Living artificial heart valve alternatives. European Cells and Materials, 6, 28-45.

Flor, H. (2002). Phantom-limb pain: characteristics, causes, and treatment. Lancet Neurology, 1(3), 182-189.

Freund, J.A. and Poschel, T. (2000). Stochastic processes in physics, chemistry, and biology. Lecture Notes in Physics, vol. 557. Springer, Berlin.

Full, R. J. and Koditschek, D. E. (1999). Templates and anchors: neuromechanical hypotheses of legged locomotion on land.   Journal of Experimental Biology, 202, 3325–3332.

Garcia, G.J.M. and Leal da Silva, J.K. (2006). Interspecific allometry of bone dimensions: A review of the theoretical models. Physics of Life Reviews, 3(3), 188-209.

Georgopoulos, A.P., Kalaska, J.F., Caminiti, R. and Massey, J.T. (1982). On the relations between the direction of two-dimensional arm movements and cell discharge in primate motor cortex. Journal of Neuroscience, 2, 1527-1537.

Gibson, J.J. (1979). Ecological approach to visual perception. Lawrence Erlbaum, Hillsdale, NJ.

Gillespie, R.B. and Sovenyi, S. (2006). Model-based cancellation of biodynamic feedthrough using a force-reflecting joystick. Journal of Dynamic Systems, Measurement, and Control, 128(1), 94-103.

Giszter, S.F., Mussa-Ivaldi, F.A., Bizzi, E. (1993). Convergent force fields organized in frog's spinal cord. Journal of Neuroscience, 13, 467-491.

Graziano, M.S.A. (2009). The Intelligent Movement Machine: an ethological perspective on the primate motor system. Oxford University Press, Oxford, UK.

Graziano, M.S.A., Aflalo, T., and Cooke, D.F. (2005). Arm movements evoked by electrical stimulation in the motor cortex of monkeys. Journal of Neurophysiology, 94, 4209-4223.

Gutierrez, A. and Zarco, G. (2003). Indicator of myocardial ischemia based on the mean power of ECG low frequency content: comparison with ST segment trend. Computers in Cardiology, 21-24, 653-656.

h2.0 Conference Webcast. (2007). h2.0 conference: new minds, new bodies, new identities. http://http://h20.media.mit.edu/ Massachusetts Institute of Technology, Cambridge, MA.

Hamburger, V., Berns, K., Iida, F, and Pfeifer, R. (2006). Standing up with motor primitives. In "Climbing and Walking Robots". Tokhi, M.O., Virk, G.S., and Hossain, M.A. eds. Springer, Berlin.

Hara, F. and Pfeifer, R. (2003). Morpho-Functional Machines: The New Species: Designing Embodied Intelligence. Springer, Berlin.

Herr, H.M., Huang, G.T., and McMahon, T.A. (2002). A model of scale effects in mammalian quadrupedal running. Journal of Experimental Biology, 205, 959-967.

Holmes, P., Full, R.J., Koditschek, D., Guckenheimer, J. (2006). The Dynamics of Legged Locomotion: Models, Analyses, and Challenges. SIAM Review, 48(2), 207-304.

Hubler, A. and Gintautas, V. (2008). Experimental Evidence for Mixed Reality States. Complexity, 13(6), 7-10.

Ishibashi, H., Hihara, S., Takahashi, M., Heikeb, T., Yokota, T., and Iriki, A. (2002). Tool-use learning selectively induces expression of brain-derived neurotrophic factor, its receptor trkB, and neurotrophin 3 in the intraparietal multisensory cortex of monkeys. Cognitive Brain Research, 14, 3–9.

Jagacinski, R.J. and Flach, J.M. (2003). Control Theory for Humans: quantitative approaches to modeling performance. CRC Press, Boca Raton, FL.

Jeka, J.J. (2006). Light touch contact: not just for surfers. The Neuromorphic Engineer, 3(1), 5-6.

Johnson, F.E. and Virgo, K.S. (2006). The Bionic Human: health promotion for people with implanted prosthetic devices. Humana Press, Totowa, NJ.

Kawato, M. (1999). Internal models for motor control and trajectory planning. Current Opinion in Neurobiology, 9, 718-727.

Koch, C. (1999). Biophysics of Computation: Information Processing in Single Neurons. Oxford University Press, New York.

Krakauer, J.W. and Shadmehr, R. (2006). Consolidation of motor memory. Trends in Neurosciences, 29, 58-64.

Krishna, S., Jensen, M.H., and Sneppen, K. (2006). Minimal model of spiky oscillations in NF-KB signaling. PNAS USA, 103(29), 10840-10845.

Lackner, J.R. and DiZio, P. (2000). Human orientation and movement control in weightlessness and artificial gravity environments. Experimental Brain Research, 130, 2-26.

Liu, J.Z., Dai, T.H., Sahgal, V., Brown, R.W. and Yue, G.H. (2002). Nonlinear cortical modulation of muscle fatigue: a functional MRI study. Brain Research, 957, 320–329.

Marr, D. (1969). A Theory of Cerebellar Cortex. Journal of Physiology, 202, 437-470.

Martin, J.T. and Nguyen, D.H. (2004). Anthropometric analysis of homosexuals and heterosexuals: implications for early hormone exposure. Hormones and Behavior, 45, 31–39.

Martin, T.A., Keating, J.G., Goodkin, H.P., Bastian, A.J., Thach, W.T. (1996). Throwing while looking through prisms. I. Focal olivocerebellar lesions impair adaptation. Brain, 119:1183–1198.

McEwen, B.S. and Stellar, E. (1993). Stress and the individual: mechanisms leading to disease. Archives of Internal Medicine, 153, 2093-2101.

McHugh, J. (2007). Blade Runner. Wired, 15.03. http://www.wired.com/wired/archive/15.03/blade.html?pg=3&topic= blade&topic_set=

McKinstry, J.L., Edelman, G.M., and Krichmar, J.L. (2006). A cerebellar model for predictive motor control tested in a brain-based device PNAS USA, 103, 3387-3392.

Mellor, J.R. (2006). Synaptic plasticity of kainate receptors. Biochemical Society Transactions, 34(5), 949-951.

Moller, A.R. (2006). Reprogramming the Brain. Progress in Brain Research, Volume 157. Elseveir Science, Amsterdam.

Morasso, P., Bottaro, A., Casadio, M., and Sanguinetti, V. (2005). Preflexes and internal models in biomimetic robot systems. Cognitive Processing, 6(1), 25-36.

Morasso, P. and Sanguinetti, V. (1994). Self-organizing topographic maps and motor planning. In "From Animal to Animats 3: proceedings of the third international conference of adaptive behavior". MIT Press, Cambridge, MA.

Mozdziak, P.E., Pulvermacher, P.M., and Schultz, E. (2001). Muscle regeneration during hindlimb unloading results in a reduction in muscle size after reloading. Journal of Applied Physiology, 91, 183-190.

Mussa-Ivaldi, F.A. (1999). Motor Primitives, Force-Fields and the Equilibrium Point Theory. In From Basic Motor Control to Functional Recovery. N. Gantchev and G. N. Gantchev eds. Academic Publishing House, Sofia.

Nicolelis, M. A. (2003). Brain-machine interfaces to restore motor function and probe neural circuits. Nature Reviews Neuroscience, 4(5), 417-422.

Nishikawa, K., Biewener, A.A., Aerts, P., Ahn, A.N., Chiel, H.J., Daley, M.A., Daniel, T.L., Full, R.J., Hale, M.E., Hedrick, T.L., Lappin, A.K., Nichols, T.R., Quinn, R.D., Satterlie, R.A., and Szymik, B. (2007). Neuromechanics: an integrative approach for understanding motor control. Integrative and Comparative Biology, 47(1), 16-54.

Norrbom, J., Sundberg, C.J., Amelin, H., Kraus, W.E., Jansson, E., and Gustafsson, T. (2004). PGC-1α mRNA expression is influenced by metabolic perturbation in exercising human skeletal muscle. Journal of Applied Physiology, 96, 189-194.

Nowak, M.A. (2006). Evolutionary Dynamics: exploring the equations of life. Harvard University Press, Cambridge. MA.

Oie, K., Kiemel, T., Jeka, J.J. (2002). Multisensory fusion: Simultaneous re-weighting of vision and touch for the control of human posture. Cognitive Brain Research, 14, 164-176.

Palacios-Pru, E. and Colasante, C. (2004). Ultrastructural reversible changes in fish neuromuscular junctions after chronic exercise. Journal of Neuroscience Research, 19(2), 245-251.

Payton, O.D., Su, S., and Meydrech, E.F. (1974). Abductor digiti quinti shuffleboard: a study in motor learning. Archives of Physical Medicine and Rehabilitation, 57(4), 169-174.

Porro, C.A., Francescato, M.P., Cettolo, V., Diamond, M.E., Baraldi, P., Zuiani, C., Bazzocchi, M., and di Prampero, P.E. (1996). Primary Motor and Sensory Cortex Activation during Motor Performance and Motor Imagery: A Functional Magnetic Resonance Imaging Study. Journal of Neuroscience, 16(23), 7688-7698.

Potter, S.M., DeMarse, T.M., Bakkum, D.J., Booth, M.C., Brumfield, J.R., Chao, Z., Madhavan, R., Passaro, P.A., Rambani, K., Shkolnik, A.C., Towal, R.B., and Wagenaar, D.A. (2004). Hybrots: hybrids of living neurons and robots for studying neural computation. Brain Inspired Cognitive Systems, ASME, Stirling, Scotland, UK.

Prochazka, A. and Yakovenko, S. (2007). Predictive and reactive tuning of the locomotor CPG. Integrative and Comparative Biology, 47(4), 474-481.

Quartarone, A., Bagnato, S., Rizzo, V., Morgante, F., SantAngelo, A., Battaglia, F., Messina, C., Siebner, H.R. and Girlanda, P. (2004). Distinct changes in cortical and spinal excitability following high-frequency repetitive TMS to the human motor cortex. Experimental Brain Research, 161(1), 114-124.

Rafiq, A., Hummel, R., Lavrentyev, V., Derry, W., Williams, D., and Merrell, R.C. (2006). Microgravity effects on fine motor skills: tying surgical knots during parabolic flight. Aviation, Space, and Environmental Medicine, 77, 852-856.

Ramirez, R.W. (1984). The FFT: fundamentals and concepts. Prentice Hall, Saddle River, NJ.

Reger, B.D., Fleming, K.M., Sanguineti, V., Alford, S., and Mussa-Ivaldi, F.A. (2000). Connecting brains to robots: an artificial body for studying the computational properties of neural tissues. Artificial Life, 6(4), 307-324.

Rice, G.M., Vacchiano, C.A., Moore, J.L., and Anderson, D.W. (2003). Incidence of decompression sickness in hypoxia training with and without 30-minute $O_2$ prebathe. Aviation, Space, and Environmental Medicine, 74, 56-61.

Ritter, A.B., Reisman, S., and Michniak, B.B. (2005). Biomedical Engineering Principles. CRC Press, Boca Raton, FL.

Roland, P.E., Larsen, B., Lassen, N.A., and Skinhoj, E. (1980). Supplementary motor area and other cortical areas in organization of voluntary movements in man. Journal of Neurophysiology, 43, 118-136.

Rome, L.C., Cook, C., Syme, D., Connaughton, M., Ashley-Ross, M., Klimov, A., Tikunov, B. and Goldman, Y.E. (1999). Trading force for speed: crossbridge kinetics of super-fast muscle fibers. PNAS USA, 96, 5826-5831.

Saper, C.B., Scamell, T.E., and Lu, J. (2005). Hypothalamic regulation of sleep and circadian rhythms. Nature, 437(7063), 1257-1263.

Selzer, M., Clarke, S., Cohen, L., Duncan, P., and Gage, F. (2006). Textbook of Neural Repair and Rehabilitation. Cambridge Press, Cambridge, UK.

Shadmehr, R. and Wise, S.P. (2005). The computational neurobiology of reaching and pointing. MIT Press, Cambridge, MA.

Shadmehr, R. and Moussavi, Z.M. (2000). Spatial generalization from learning dynamics of reaching movements. Journal of Neuroscience, 20, 7807-7815.

Shadmehr, R. and Brashers-Krug, T. (1997) Functional stages in the formation of human long-term motor memory. Journal of Neuroscience, 17: 409-419.

Shadmehr, R. and Holcomb, H.H. (1997). Neural correlates of motor memory consolidation. Science, 277(5327), 821-825.

Shadmehr, R., and Mussa-Ivaldi, F.A. (1994). Rapid adaptation to Coriolis force perturbations of arm trajectory. Journal of Neuroscience, 14, 3208-3224.

Sheridan, T.B. (1993). Space teleoperation through time delay. IEEE Transactions on Robotics and Automation, 9(5), 592-606.

Smith, D.H., Wolf, J.A., and Meaney, D.F. (2001). A New Strategy to Produce Sustained Growth of Central Nervous System Axons: Continuous Mechanical Tension. Tissue Engineering, 7(2), 131-139.

Snyder, J.B., Nelson, M.E., Burdick, J.W., MacIver, M.A. (2007). Omnidirectional Sensory and Motor Volumes in Electric Fish. PLoS Biology, 5(11), e301

Sterling, P. (2004). Principles of allostasis: optimal design, predictive regulation, pathophysiology and rational therapeutics. IN Allostasis, Homeostasis, and the Costs of Adaptation. J. Schulkin ed.Cambridge University Press, Oxford, UK.

Sterling, P. and Eyer, J. (1988). Allostasis: a new paradigm to explain arousal pathology. Handbook of Life Stress, Cognition, and Health, S. Fisher and J. Reason eds., Wiley and Sons, New York.

Storlien, L., Oakes, N.D., and Kelley, D.E. (2004). Metabolic Flexibility. Symposium 6: Adipose tissue–liver–muscle interactions leading to insulin resistance. Proceedings of the Nutrition Society, 63, 363–368.

Sweatt, J.D. (2003). Mechanisms of Memory. Elsevier, Amsterdam.

Tin, C. and Poon, C-S. (2005) Internal Models in Sensiorimotor Integration: perspectives from adaptive control theory. Journal of Neural Engineering, 2, S147-S163.

von Holst, E. and Mittelstaedt, H. (1980). The reafference principle: interaction between the central nervous system and the periphery. In "The Organization of Action", C.R. Gallistel ed., Lawrence Erlbaum, Hillsdale, NJ.

Winter, D.A. (1990). Biomechanics and Motor Control of Human Movement. Wiley, New York.

Wissler, E.H. (2003). Probability of survival during accidental immersion in cold water. Aviation, Space, and Environmental Medicine, 74, 47-55.

Wolfe, J.M., Kluender, K.R., Levi, D.M., Bartoshuk, L.M., Herz, R.S., Klatzky, R.S., Lederman, S.J. (2005). Sensation and Perception. Sinauer Associates, Sunderland, MA.

Wolpaw, J.R. and Tennissen, A.M. (2001). Activity-dependent spinal cord plasticity in health and disease. Annual Review of Neuroscience, 24, 807–843.

Wolpert, D.M., Ghahramani, Z., and Jordan, M.I. (1995). An internal model for sensorimotor integration. Science, 269, 1880-1882.

Wu, Z., Puigserver, P., Andersson, U., Zhang, C., Adelmant, G., Mootha, V., Troy, A., Cinti, S., Lowell, B., Scarpulla, R.C., and Speigelman, B.M. Mechanisms controlling mitochondrial biogenesis and respiration through the thermogenic coactivator PGC-1, Cell 98, 115-124 (1999).